\documentclass[twocolumn,trackchanges,sort&compress,numbers,twocolappendix]{aastex631}
\graphicspath{{./}{figures/}}

\begin{document}

\title{Fragmentation of massive clumps: constant mass fraction and growing mass inequality of cores}
\author[0000-0002-9138-5940]{Zu-Jia Lu}\thanks{luzujia@gxu.edu.cn}
\affiliation{Guangxi Key Laboratory for Relativistic Astrophysics, Department of Physics, Guangxi University, Nanning 530004, China}
\affiliation{Department of Physics, University of Oxford, Keble Road, Oxford OX1 3RH, UK}

\author{Qi-Tian Chen}
\affiliation{Guangxi Key Laboratory for Relativistic Astrophysics, Department of Physics, Guangxi University, Nanning 530004, China}




\begin{abstract}

The fragmentation of massive molecular clumps into dense cores sets the initial conditions for stellar clusters, yet how core mass distributions evolve remains unclear. Using the ALMAGAL survey (1,013 clumps and 6,348 cores), we analyze the Gini coefficient of core mass distributions for the 514 clumps containing at least four cores ($N_{\rm frag} \ge 4$), which host 5,728 cores in total. We find that the most massive core consistently accounts for $\sim 38\%$ of the total core mass ($f_{\rm core,max} \approx 0.38$), with no systematic trend with the total mass of the core per clump $M_{\rm core,tot}$. The Gini coefficient increases with total core mass ($M_{\rm core,tot}$, $r=0.55$) and evolutionary stage ($L/M$, $r=0.38$). Clumps with $L/M > 6.0$ have significantly higher Gini values (median $0.62$) than those with $L/M \le 6.0$ (median $0.49$; $p = 4.23\times10^{-15}$). These results demonstrate that while the mass fraction of the most massive core is constant, overall mass inequality among cores grows systematically as clumps evolve.

\end{abstract}




\section{Introduction} \label{introduction}

The formation of stars and star clusters is a hierarchical process that spans scales from giant molecular clouds ($\sim 10-100$~pc) to dense cores ($\sim 10^3$~au). A critical intermediate step is the fragmentation of massive molecular clumps ($\sim 0.1-1$~pc) into populations of compact dense cores, which directly sets the initial conditions for stellar clusters \citep{Bonnell+2004MNRAS, Tan+2014PPV, Krumholz+2019ARA&A}. The resulting core multiplicity - the number and mass distribution of cores within a clump - determines whether stars form in isolation or in groups, and influences the final stellar initial mass function (IMF).

The ALMA evolutionary study of high-mass protocluster formation in the Galaxy (ALMAGAL) survey has provided an unprecedented census of dense molecular cores within massive clumps across the Milky Way \citep{ALMAGAL+I+Molinari+2025A&A, ALMAGAL+III+Coletta+2025A&A}. With 1,013 clumps and 6,348 cores observed at $\sim 1,000$~au resolution, ALMAGAL has enabled detailed studies of core mass functions \citep{ALMAGAL+III+Coletta+2025A&A}, spatial distributions \citep{ALMAGAL+VI+Schisano+2026A&A}, and relations between core populations and clump properties \citep{ALMAGAL+V+Elia+2026A&A}.

A central question in star formation is whether the final stellar masses are determined at the core formation epoch, or whether significant mass growth occurs after cores emerge. Observations in low-mass star-forming regions (e.g., $\rho$ Ophiuchi) have shown that the core mass function (CMF) closely resembles the IMF, suggesting a direct mapping \citep{Motte+1998A&A, Alves+2007A&A}. However, studies of high-mass star-forming regions have revealed a more complex picture, with growing evidence that cores continue to accrete mass after fragmentation. The ALMAGAL survey found that the CMF slope evolves with clump evolutionary stage, becoming shallower at later stages, indicating ongoing mass growth \citep{ALMAGAL+III+Coletta+2025A&A}. Indeed, \citet{ALMAGAL+III+Coletta+2025A&A} showed that the most massive cores grow in mass with evolution while a population of low-mass cores persists, suggesting a clump-fed scenario in which cores form as low-mass seeds and then gain mass from the surrounding intraclump medium. Similarly, \citet{Xu+2024ApJS} found that cores in evolved massive clumps (ASSEMBLE survey) have average masses about 2-3 times higher than those in earlier stage IRDCs (ASHES survey), providing further evidence for core mass growth with evolution. Numerical simulations also support this picture: \citet{Pelkonen+2021MNRAS} show that there is only a weak correlation between the progenitor core mass and the final stellar mass for any individual star, implying that significant mass redistribution occurs after core formation. Taken together, these findings suggest that cores are not static entities with predetermined masses; instead, they grow dynamically by accreting mass from their surroundings after fragmentation.

While these studies have established fundamental correlations, they have primarily focused on average properties of core populations. Less attention has been paid to the \textit{internal structure} of core populations - specifically, how mass is distributed among cores within a single clump. Is the mass budget dominated by a single massive core, or is it more evenly distributed among many low-mass cores? Does this distribution change as clumps evolve? Answering these questions is essential for distinguishing between competing models of massive star formation. The competitive accretion scenario \citep{Bonnell+2001MNRAS} predicts that cores grow by accreting from a shared reservoir, leading to increasingly unequal mass distributions over time. In contrast, monolithic collapse models \citep{McKee+2003ApJ} assume core masses are predetermined at formation, predicting no systematic evolution in mass inequality. Quantifying the internal mass distribution of core populations thus provides a direct test of these competing predictions.

In this paper, we study these questions using two complementary metrics: the mass fraction of the most massive core ($f_{\rm core,max} \equiv M_{\rm core,max}/M_{\rm core,tot}$) and the Gini coefficient of the core mass distribution. The Gini coefficient, widely used in astronomy to quantify inequality \citep{Goyal+2022ApJ}, captures the full shape of the mass distribution rather than focusing only on the most extreme core. Together, these two metrics provide complementary information: $f_{\rm core,max}$ tracks the dominance of the single most massive core, while the Gini coefficient quantifies the overall inequality across the entire core population. We apply these metrics to ALMAGAL clumps with at least four cores ($N_{\rm frag} \ge 4$), a sample of 514 clumps containing 5,728 cores, investigating how core mass inequality correlates with total core mass ($M_{\rm core,tot}$) and clump evolutionary stage ($L/M$).

While the ALMAGAL survey and the properties of the massive clumps have been described in detail in \citet{ALMAGAL+I+Molinari+2025A&A} (Paper I) and the dense molecular core catalog presented in \citet{ALMAGAL+III+Coletta+2025A&A} (Paper III), the analysis presented in this paper focuses on two aspects that have not been explicitly quantified in the ALMAGAL series: the constancy of the mass fraction of the most massive core ($f_{\mathrm{core,max}}$) and the evolution of core mass inequality within the entire core population, as quantified by the Gini coefficient. Paper III extensively studied the mass growth of the most massive cores and their physical properties (see their Section~7 and Figure~19), and the right panel of their Figure~19 hints at a nearly constant $f_{\mathrm{core,max}}$. We further explicitly investigate this constancy and a broader question of how mass is distributed among all cores within a clump, i.e., whether the mass budget is dominated by a single core or shared more equally among multiple cores. Our work addresses this gap by examining the overall mass inequality of core populations and its dependence on clump properties and evolution.

This paper is organized as follows. Section~\ref{data} describes the ALMAGAL data and methods. Section~\ref{results} presents the main results on $f_{\rm core,max}$ and the Gini coefficient. Section~\ref{discussion} discusses the implications and Section~\ref{summary} summarizes our main findings.

\section{Data and Methods}
\label{data}

%
\begin{figure*} 
\centering  
\includegraphics[width=0.49\linewidth]{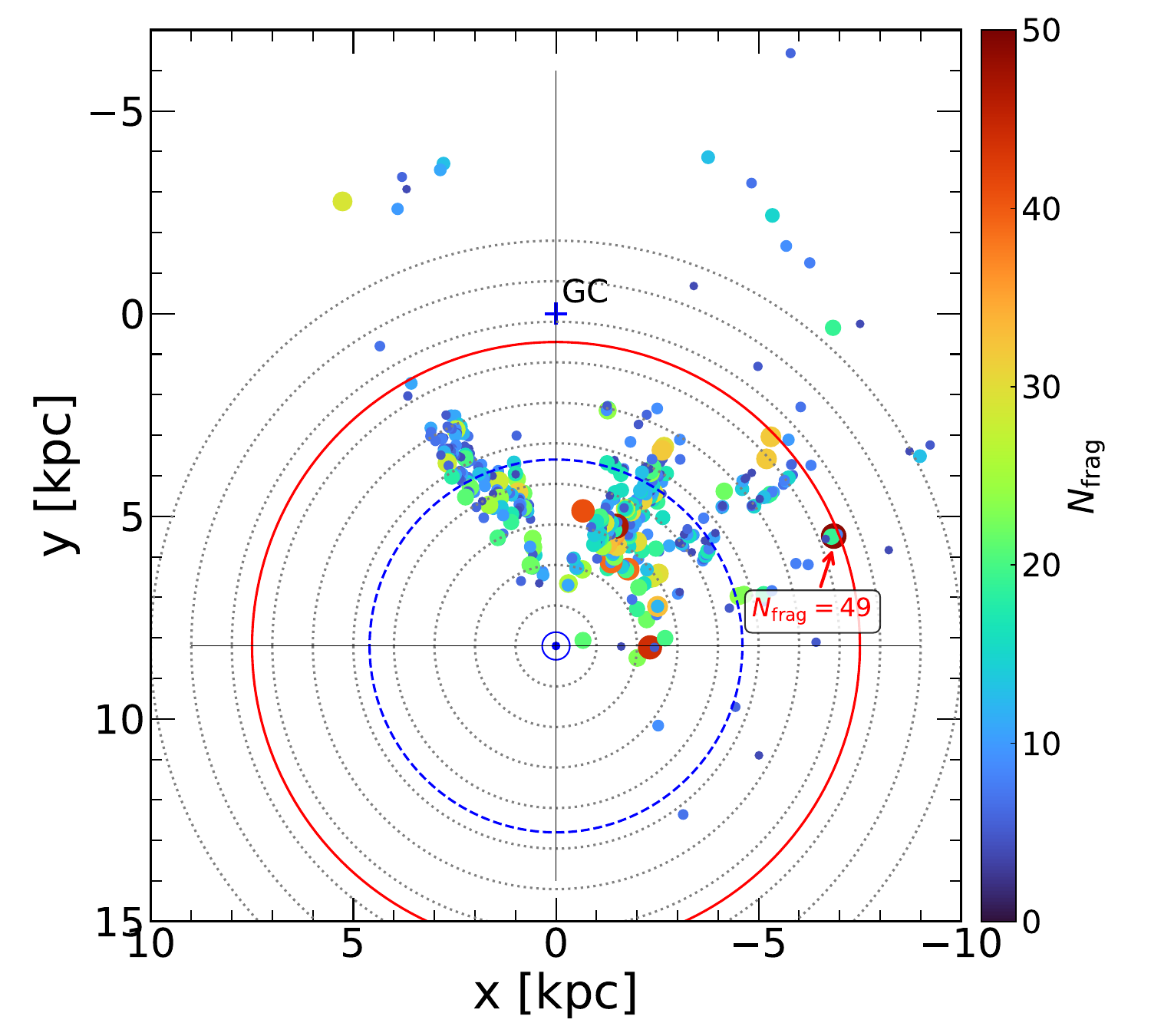}
\includegraphics[width=0.49\linewidth]{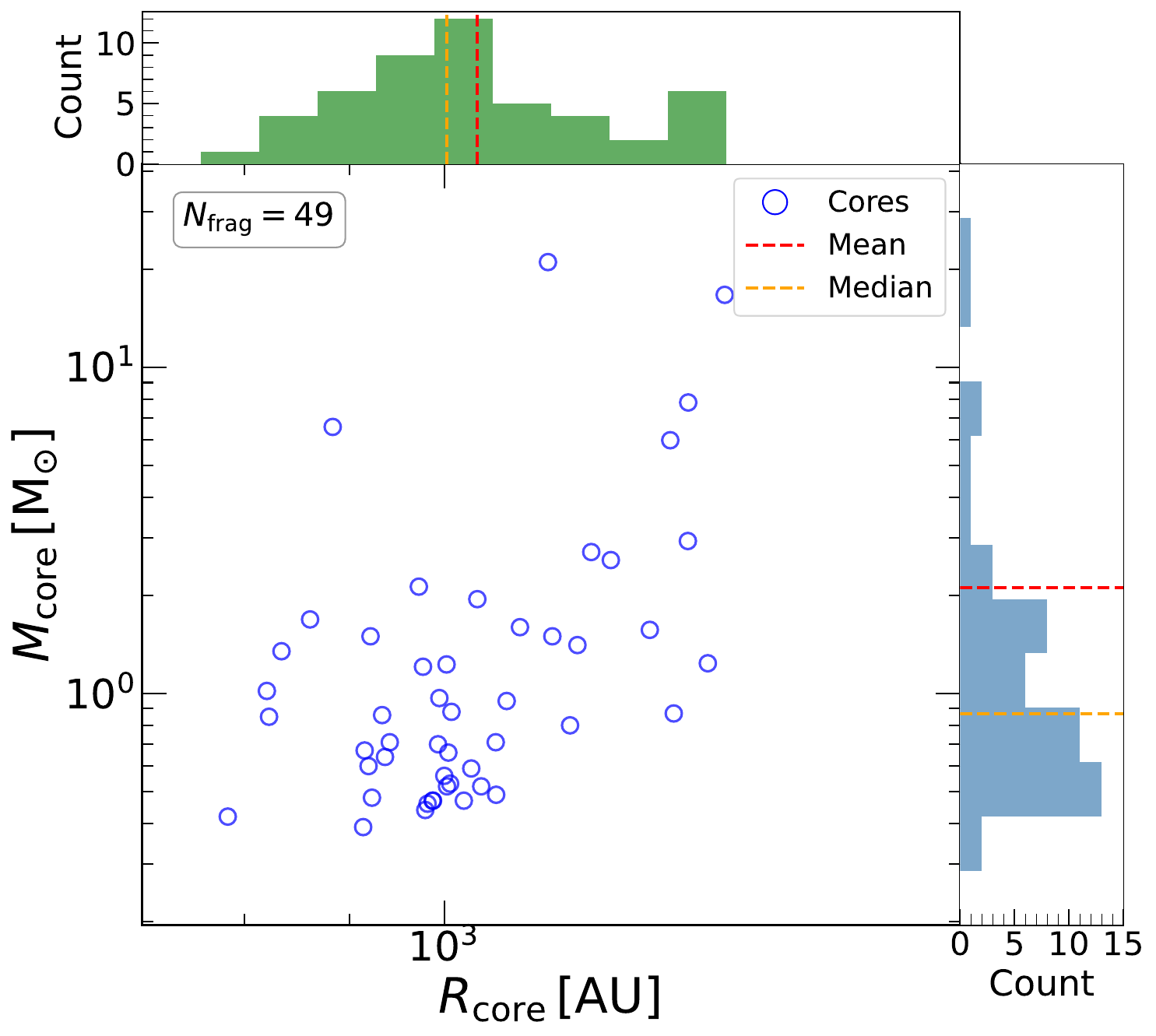}
\caption{\textit{Left panel}: Galactic distribution of the ALMAGAL massive clumps. The symbol size scales with the number of cores per clump, and the color coding also represents $N_{\rm frag}$. The statistics are derived from the ALMAGAL paper III catalog (Table~3 in \citet{ALMAGAL+III+Coletta+2025A&A}). Dashed circles indicate distance intervals of 1~kpc centered on the Sun, with the cross marking the Galactic center. The blue dashed circle marks the distance threshold used to assign the targets to the two ALMA array configurations, designed to provide a linear resolution of at least 1,000~au. The red circle indicates the $7.5$~kpc distance initially used as the upper limit for source selection.
\textit{Right panel}: Correlation of the core mass and radius in the most fragmented clump, with $N_{\rm frag}$ = 49. The distributions of the mass and radius are in the subpanels, with red and yellow dashed lines marking the mean and median values, respectively.}
\label{fig_galactic_distribution}  
\end{figure*}
%

\subsection{ALMAGAL massive clump and dense core catalogs}

The data used in this work are from the ALMAGAL survey, an ALMA Cycle 7 Large Program (2019.1.00195.L, PIs: S. Molinari, P. Schilke, C. Battersby, and P. Ho) designed to study the fragmentation of massive clumps across the Milky Way \citep{ALMAGAL+I+Molinari+2025A&A}. ALMAGAL observed 1,013 massive clumps at 1.38~mm with a single-pointing strategy targeting the $250\ \mu\mathrm{m}$ peaks from Herschel Hi-GAL \citep{Molinari+2016bA&A,Elia+2017A&A}. Using both the ACA $7\ \mathrm{m}$ and main $12\ \mathrm{m}$ arrays with two configurations, the survey achieves a median spatial resolution of $\sim 1,400\ \mathrm{au}$ (ranging from $\sim 800$ to $2,000\ \mathrm{au}$), sufficient to resolve typical high-mass star-forming cores \citep{SanchezMonge+2013A&A,Beuther+2018A&A,Sanhueza+2019ApJ}.

The ALMAGAL clump catalog contains 1,013 massive clumps selected from the Herschel Infrared Galactic Plane Survey (Hi-GAL) \citep{Molinari+2010A&A,Elia+2017A&A,Elia+2021A&A} and the Red Midcourse Space Experiment Source (RMS) \citep{Lumsden+2013ApJS} surveys. Clump masses range from $\sim 100$ to $12,000\,\rm M_{\odot}$, bolometric luminosities from $\sim 50$ to $5\times10^5\,\rm L_{\odot}$, and $L/M$ ratios from $0.05$ to $450\,\rm L_{\odot}/M_{\odot}$, covering the full evolutionary sequence from infrared-dark clouds to HII regions \citep{Molinari+2016aApJ,Traficante+2023MNRAS}. Surface densities range from $\sim 0.1$ to $10\,\rm g\,cm^{-2}$, consistent with thresholds for high-mass star formation \citep{Kauffmann+2010ApJL,Tan+2014PPV}. 

Compact cores were extracted from the 1.38~mm images using a modified CuTEx algorithm \citep{ALMAGAL+III+Coletta+2025A&A}, yielding 6,348 cores detected at $\geq 5\sigma$ in 844 clumps (83\% of the sample). The fragmentation level ranges from 1 to 49 cores per clump (median 5). Core masses were derived assuming optically thin emission, with core temperatures assigned based on the host clump's $L/M$ ratio, ranging from 20 to $81\,\rm K$ (median $35\,\rm K$). The resulting core masses span $0.002$ to $345\,\rm M_{\odot}$ (median $0.4\,\rm M_{\odot}$), with a 90\% completeness limit at $0.23\,\rm M_{\odot}$. We note that free-free continuum emission from HII regions can contribute to the 1.38~mm flux, particularly for the most massive cores in evolved clumps. As discussed in \citet{ALMAGAL+III+Coletta+2025A&A}, this contamination was not subtracted in the core mass derivation (see Section~\ref{sec_discussion_f38} for a discussion of its impact on our results).

For our analysis, we focus on clumps with $N_{\rm frag} \ge 4$ to ensure sufficient statistics for the Gini coefficient. This yields a sample of 514 clumps hosting 5,728 cores in total. For the Gini vs $L/M$ analysis, we further require reliable clump mass measurements to compute the $L/M$ ratio, resulting in a sample of 512 clumps (from the original 514 clumps with $N_{\rm frag} \ge 4$)\footnote{Two clumps (AG014.9963-0.6733 and AG305.2021+0.2073) are excluded from the Gini vs $L/M$ analysis due to missing mass measurements.}. For the details of the massive clump and dense core catalogs, we refer the reader to ALMAGAL papers \citep{ALMAGAL+I+Molinari+2025A&A,ALMAGAL+III+Coletta+2025A&A}.

In Figure~\ref{fig_galactic_distribution} we provide an overview of the clump sample used in this work. The left panel shows the Galactic distribution of the 514 clumps with $N_{\rm frag} \ge 4$, color-coded by the number of cores per clump, with symbol size proportional to $N_{\rm frag}$, to better visualize the distribution of clump
fragmentation in Galactic plane. The right panel illustrates the mass-radius relation for the most fragmented clump in the sample ($N_{\rm frag}=49$), demonstrating the quality of the core catalog.

\begin{figure*}
\centering
\includegraphics[width=0.49\linewidth]{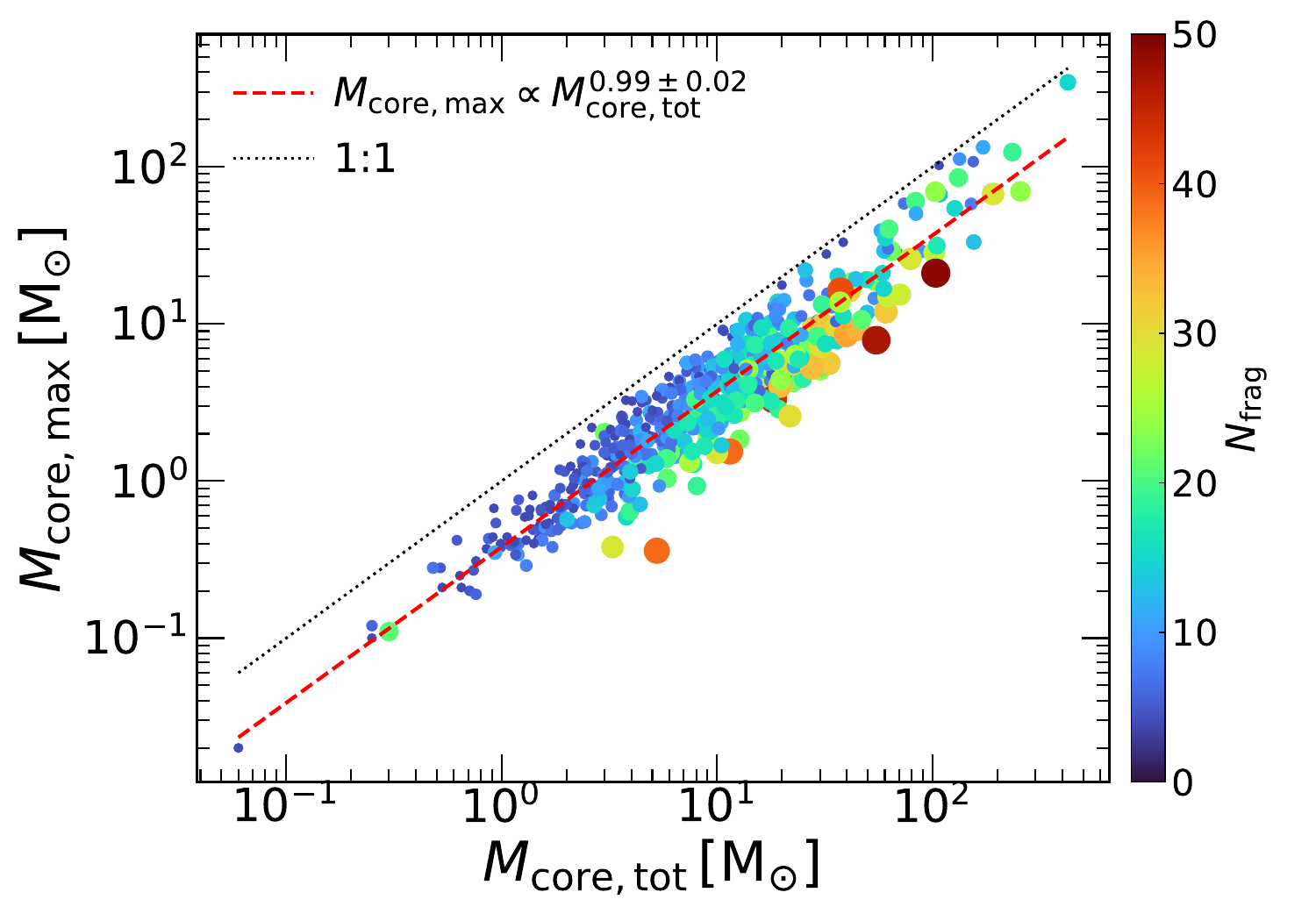}
\includegraphics[width=0.49\linewidth]{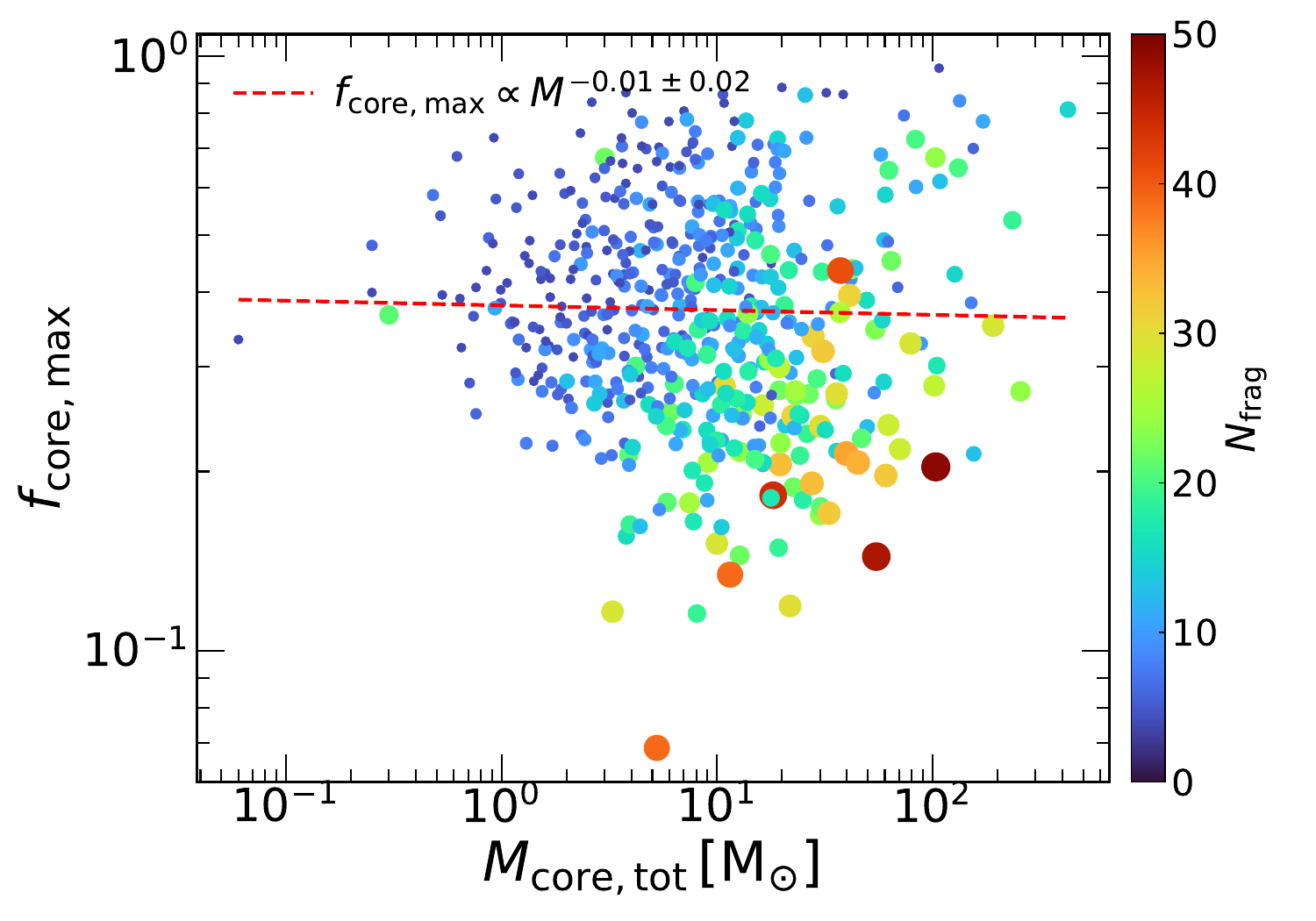}
\caption{\textit{Left panel}: mass of the most massive core ($M_{\rm core,max}$) as a function of the total core mass ($M_{\rm core,tot}$) within each ALMAGAL clump. Points are color-coded and sized by the number of cores per clump ($N_{\rm frag}$). The red dashed line shows the linear least-squares fit, and the black dotted line indicates the one-to-one relation. Pearson's $r = 0.95$. \textit{Right panel}: mass fraction of the most massive core ($f_{\rm core,max} \equiv M_{\rm core,max}/M_{\rm core,tot}$) versus $M_{\rm core,tot}$. Pearson's $r = -0.02$.}
\label{fig_max_core}
\end{figure*}

\subsection{Gini coefficient}

To quantify the inequality of core mass distributions, we adopt the Gini coefficient, a statistical measure originally developed in econometrics to assess income or wealth inequality \citep{Gini+1912amu.book}. The Gini coefficient has been increasingly applied in astrophysical contexts, including the characterization of the spatial distribution of galaxies, the uniformity of exoplanet populations, and the morphological classification of galaxy mergers \citep{Goyal+2022ApJ}. Its value ranges from 0 (perfect equality, all elements have identical values) to 1 (perfect inequality, a single element dominates the total). In this work, we use the Gini coefficient to measure the degree of mass inequality among cores within each ALMAGAL clump.

We compute two metrics for each clump: (i) the mass fraction of the most massive core $f_{\rm core,max} \equiv M_{\rm core,max}/M_{\rm core,tot}$, and (ii) the Gini coefficient of the core mass distribution, which is defined as
\begin{equation}
G_{\rm def} = \frac{\sum_{i=1}^{n} (2i - n - 1) \cdot m_{(i)}}{\sum_{i=1}^{n} (n - 1) \cdot m_{(i)}} \, ,
\end{equation}
where $m_{(i)}$ are the core masses sorted in ascending order ($m_{(1)} \le m_{(2)} \le \dots \le m_{(n)}$). The Gini coefficient ranges from 0 (perfect equality, all cores have identical masses) to 1 (perfect inequality, a single core dominates the total mass). Since the raw Gini coefficient is known to be biased for small samples ($n \lesssim 10$) \citep{Deltas2003, Goyal+2022ApJ}, we apply the standard correction,
\begin{equation}
G = G_{\rm def} \times \frac{n}{n - 1} \, .
\end{equation}
All Gini values reported in this paper are the corrected values $G$ unless otherwise noted.

\subsection{Statistical analysis}

Throughout this paper, we adopt a significance threshold of $p < 0.05$ 
for statistical significance. Correlation analyses use the Pearson 
linear correlation coefficient $r$, with $p$-values testing the null 
hypothesis of no correlation. For two-group comparisons (e.g., low 
versus high $L/M$), we use the two-sided Mann-Whitney U test, which is 
nonparametric and robust against outliers. Power-law fits are performed using ordinary least squares in logarithmic space. All reported uncertainties correspond to $1\sigma$ confidence intervals derived from the covariance matrix of the fit.

\begin{figure*}
\centering
\includegraphics[width=0.49\linewidth]{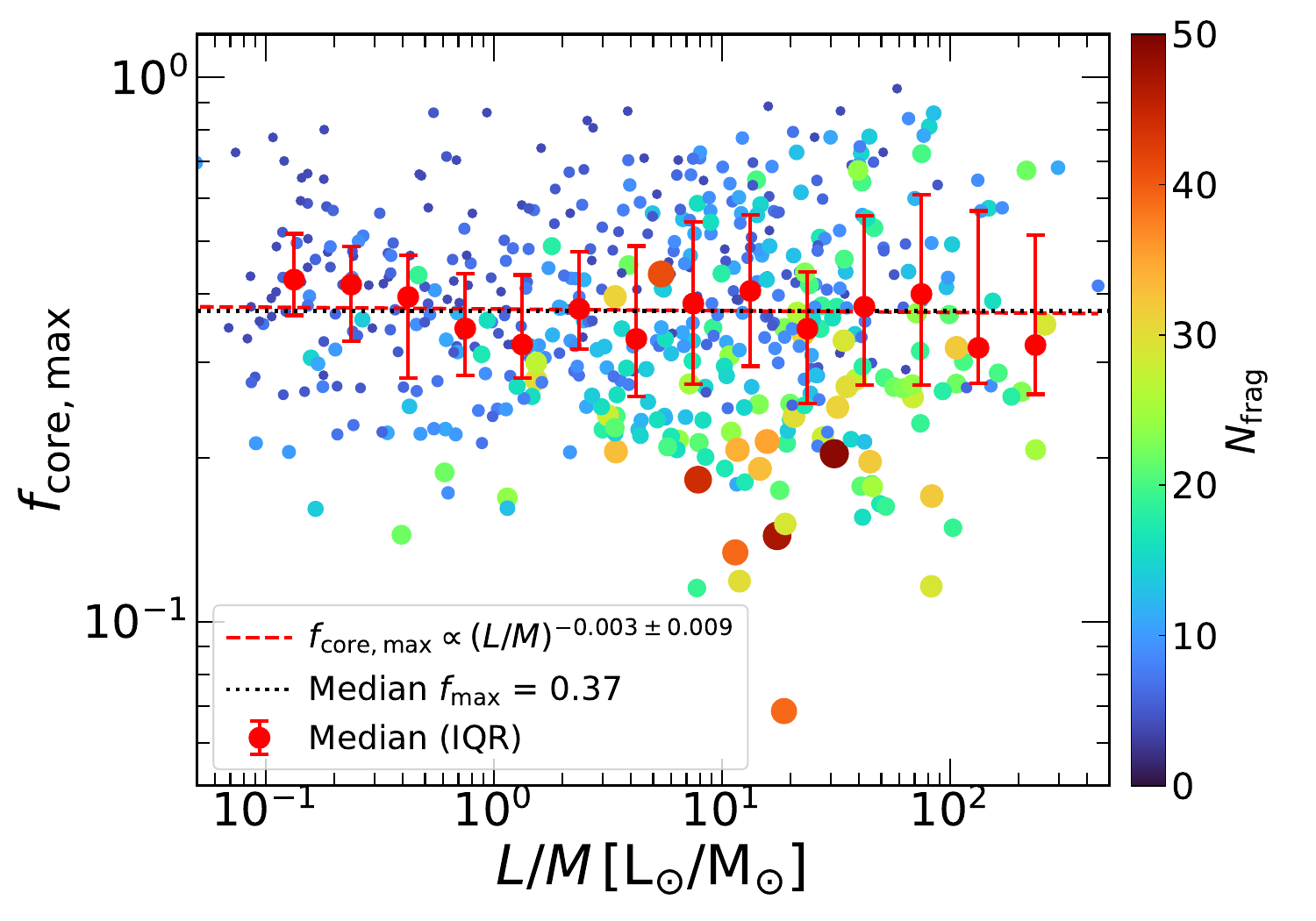}
\includegraphics[width=0.49\linewidth]{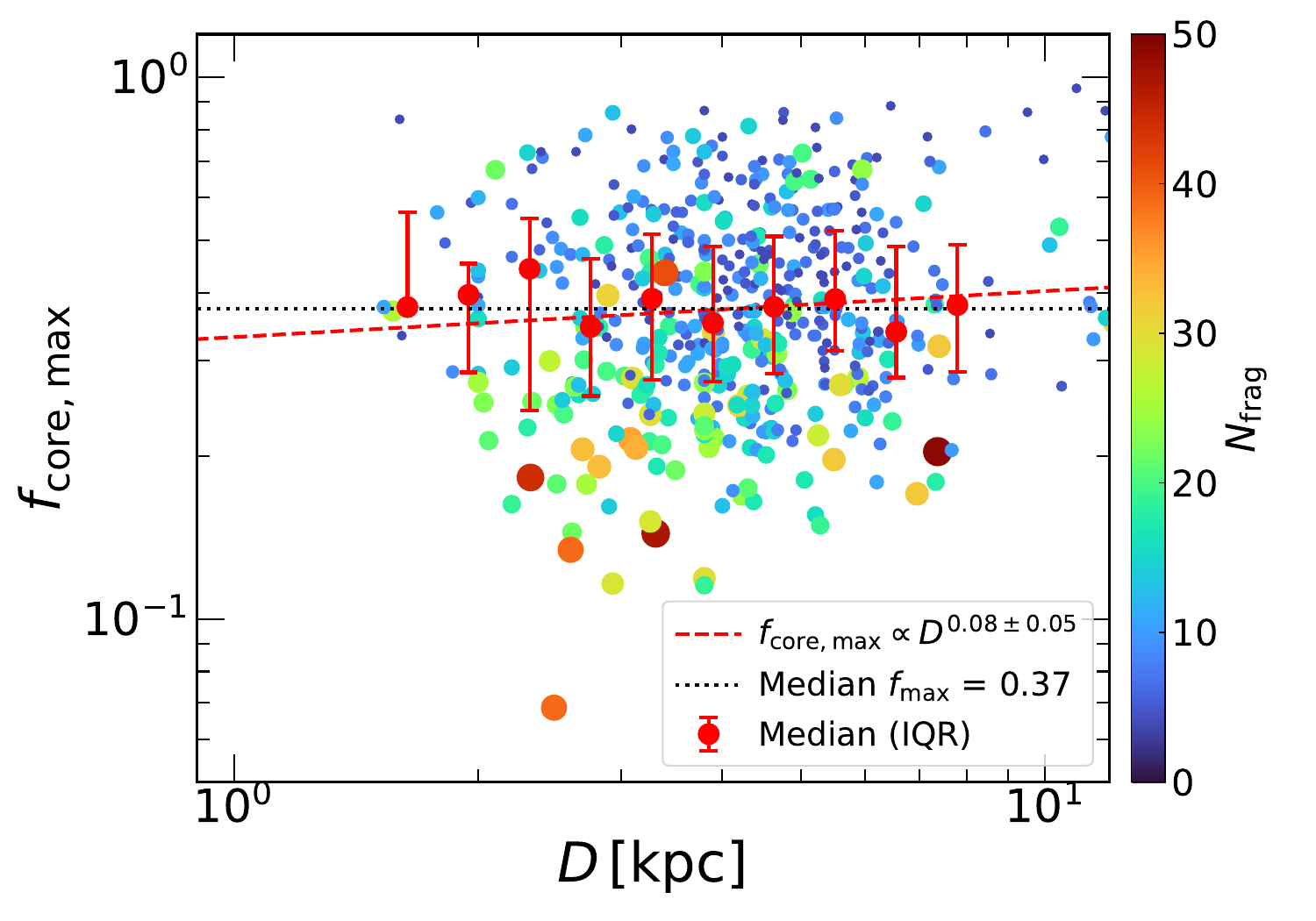}
\caption{\textit{Left panel}: $f_{\rm core,max}$ as a function of $L/M$. No significant correlation is found (Pearson $r = -0.015$, Spearman $\rho = -0.003$, $p=0.94$), indicating that the scatter is not driven by evolutionary differences. \textit{Right panel}: $f_{\rm core,max}$ as a function of source distance $D$. No significant correlation is found (Pearson $r = 0.078$, Spearman $\rho = 0.067$, $p=0.13$), ruling out distance-dependent biases. The red dashed lines show power-law fits, and the black dotted lines mark the median value of $f_{\rm core,max} = 0.37$. Points are color-coded and sized by $N_{\rm frag}$. Red circles with error bars represent the median and interquartile range (IQR; 25th-75th percentiles) in bins.}
\label{fig_max_core_lm_distance}
\end{figure*}

\section{Results}
\label{results}

\subsection{The most massive core in a clump}
\label{max_caore}

The left panel of Figure~\ref{fig_max_core} shows the correlation between the mass of the most massive core ($M_{\rm core,max}$) and the total mass of all cores ($M_{\rm core,tot}$) within each clump. The two quantities follow a tight, nearly linear relation. The red dashed line is the linear least-squares fit,
\begin{equation}
M_{\rm core,max} = 3.81^{+0.14}_{-0.13} \times 10^{-1} \, \left( \frac{M_{\rm core,tot}}{1 \,\rm M_{\rm \odot}} \right)^{0.99 \pm 0.02} \, \rm M_{\rm \odot} \, .
\label{eq_max_core}
\end{equation}
Pearson's $r$ coefficient is $0.95$. The slope is indistinguishable from unity, indicating that the most massive core consistently accounts for a nearly constant fraction of $\sim 38\%$ of the total core mass across approximately four orders of magnitude, ranging from $\sim 0.1$ to $\sim 300$ M$\odot$.

This tight correlation, with its near-unity slope, is consistent with the findings of \citet{ALMAGAL+III+Coletta+2025A&A}, which examined the relation between the most massive core mass and the total core mass in the full ALMAGAL sample (their Figure~19, right panel). Our analysis, restricted to clumps with $N_{\rm frag}\ge4$ for statistical robustness, recovers the same fundamental relation and further quantifies the constant mass fraction. While Paper III focused on the mass growth history of individual dominant cores and their physical properties, here we extend the analysis to ask: (i) whether this constant fraction is universal across clumps of different total core mass, and (ii) more importantly, whether the distribution of mass among the entire population of cores evolves with clump properties.

The right panel of Figure~\ref{fig_max_core} addresses the first question by showing the mass fraction of the most massive core ($f_{\rm core,max} \equiv M_{\rm core,max}/M_{\rm core,tot}$) as a function of $M_{\rm core,tot}$. $f_{\rm core,max}$ shows no systematic correlation with $M_{\rm core,tot}$ (Pearson's $r = -0.02$). Even for highly fragmented clumps ($N_{\rm frag} > 20$), $f_{\rm core,max}$ remains in the range $0.2$--$0.4$, suggesting a universal mass concentration efficiency in clump-fed core accretion that is remarkably independent of the total mass budget.

To investigate the origin of the scatter in $f_{\rm core,max}$ shown in the right panel of Figure~\ref{fig_max_core}, we examine its dependence on clump evolutionary stage ($L/M$) and heliocentric distance. Figure~\ref{fig_max_core_lm_distance} presents these correlations. The left panel shows $f_{\rm core,max}$ as a function of $L/M$ for the 512 clumps with valid $L/M$ measurements. No significant correlation is observed (Pearson $r = -0.015$, Spearman $\rho = -0.003$, $p = 0.94$), indicating that the scatter in $f_{\rm core,max}$ is not primarily driven by evolutionary differences among clumps. The median $f_{\rm core,max}$ values remain $\sim 0.35$-$0.40$ across the entire range of $L/M$ from $\sim 0.1$ to $\sim 450$ $\rm L_{\odot}/M_{\odot}$. The right panel of Figure~\ref{fig_max_core_lm_distance} shows $f_{\rm core,max}$ as a function of source distance for the 514 clumps with valid distance measurements. Again, no significant correlation is found (Pearson $r = 0.078$, Spearman $\rho = 0.067$, $p = 0.13$), ruling out distance-dependent detection biases as the primary source of scatter. The power-law fits yield $f_{\rm core,max} \propto (L/M)^{-0.003 \pm 0.009}$ and $f_{\rm core,max} \propto D^{0.08 \pm 0.05}$, respectively, both consistent with no dependence. The lack of correlation with both $L/M$ and distance suggests that the observed scatter in $f_{\rm core,max}$ is intrinsic to the fragmentation process rather than being driven by evolutionary stage or observational selection effects. This intrinsic scatter, however, is centered around a well-defined median value of $\sim 0.37$, further reinforcing the universality of the mass concentration efficiency.

The distribution of $f_{\rm core,max}$ across the sample is shown in Appendix~\ref{app_fmax}. We now turn to the Gini coefficient to investigate whether the overall mass inequality among all cores, beyond just the most massive one, also shows systematic trends.

\begin{figure*}
\centering
\includegraphics[width=0.49\linewidth]{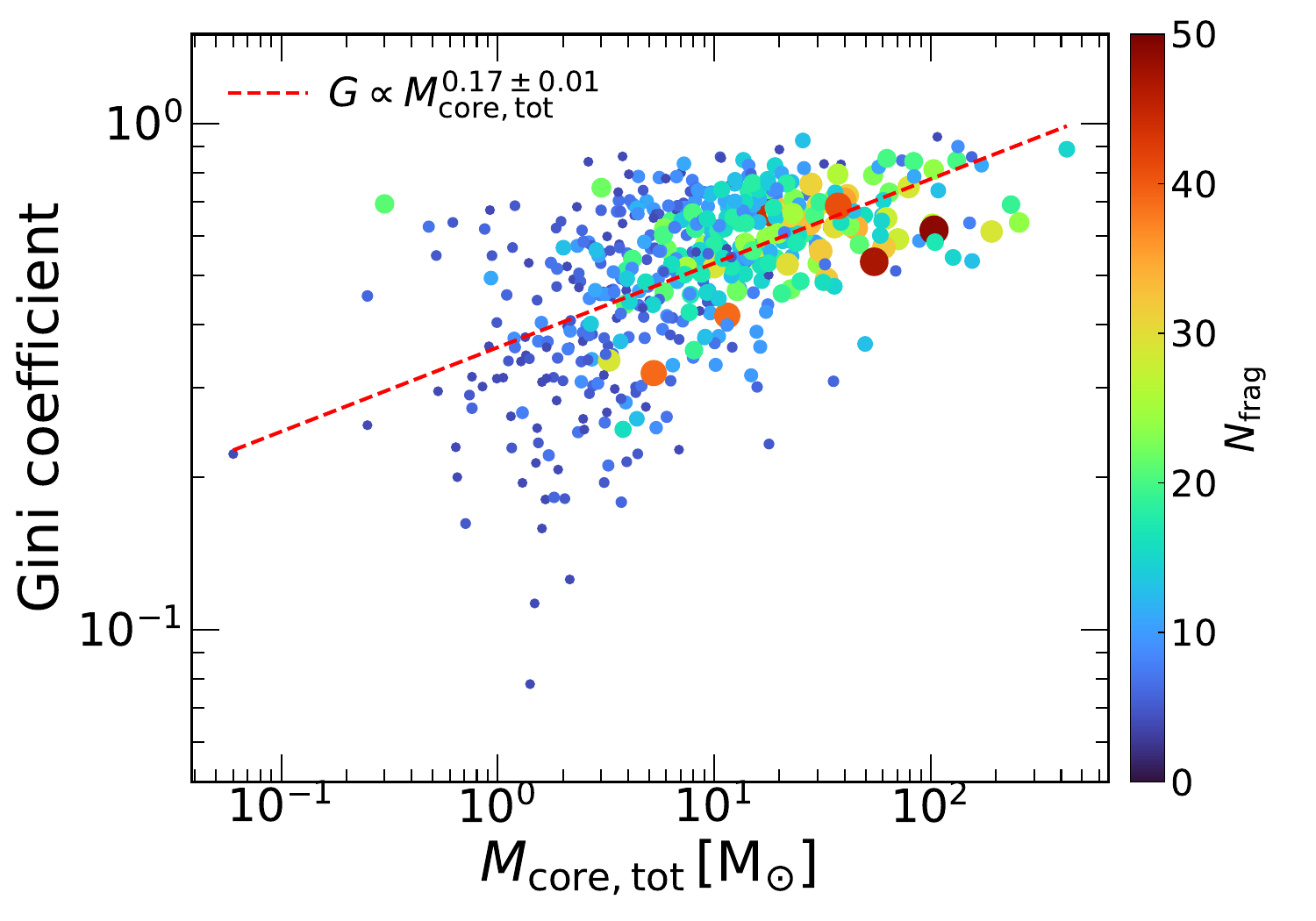}
\includegraphics[width=0.49\linewidth]{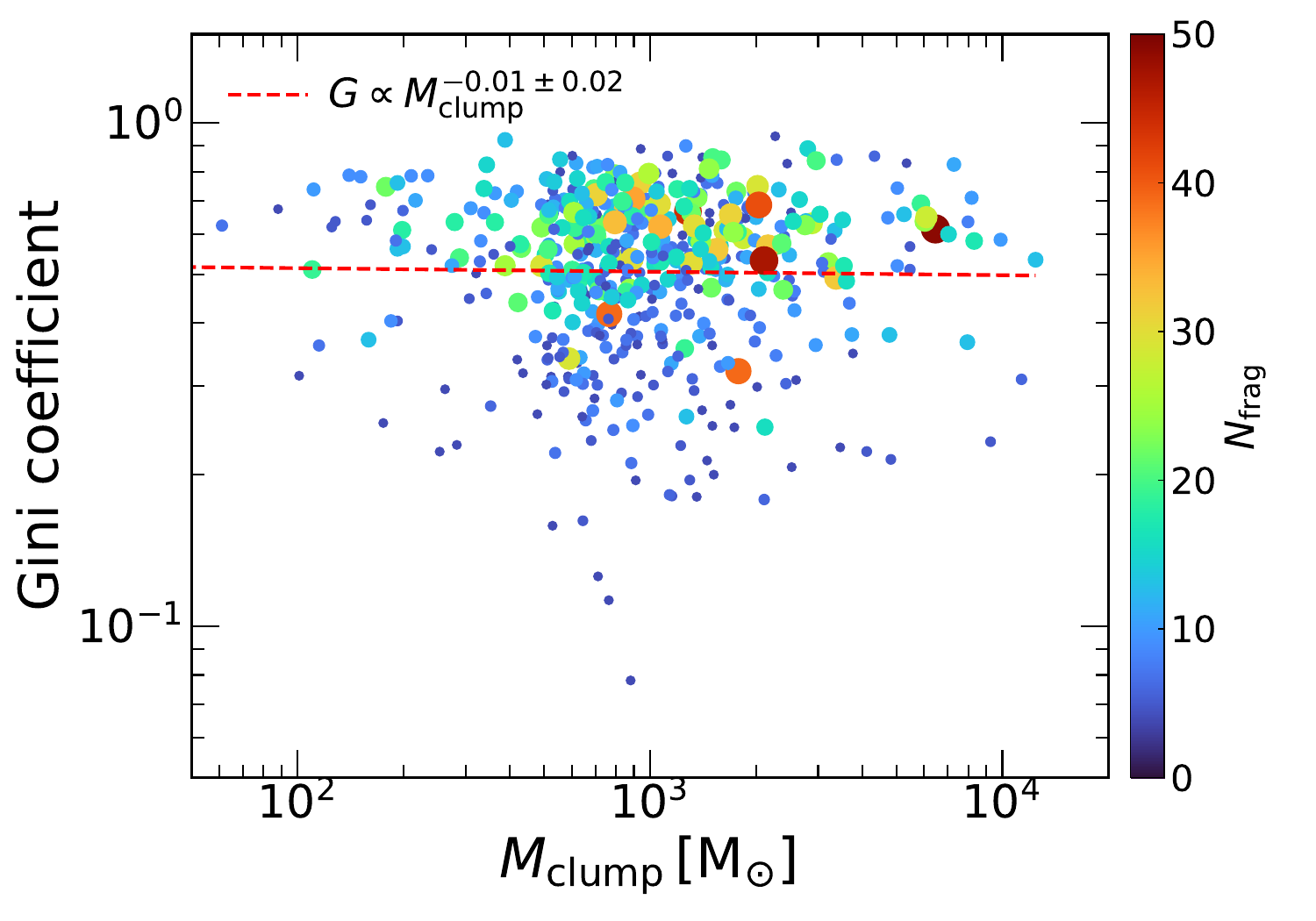}
\caption{\textit{Left panel}: Gini coefficient $G$ as a function of total core mass $M_{\rm core,tot}$. Points are color-coded and sized by the number of cores per clump ($N_{\rm frag}$). The red dashed line shows the power-law fit $ G \propto M_{\rm core,tot}^{0.17 \pm 0.01}$. Pearson's $r$ coefficient is $0.55$. \textit{Right panel}: Gini coefficient $G$ as a function of clump mass $M_{\rm clump}$. No significant correlation is found (Pearson $r = -0.016$, Spearman $\rho \approx 0.0004$, $p = 0.99$), with a power-law fit $G \propto M_{\rm clump}^{-0.01 \pm 0.02}$ consistent with no dependence. The lack of correlation with $M_{\rm clump}$ suggests that the total mass in cores, rather than the total clump mass, is the more direct driver of mass inequality.}
\label{fig_gini_mass}
\end{figure*}

\begin{figure}
\centering
\includegraphics[width=\linewidth]{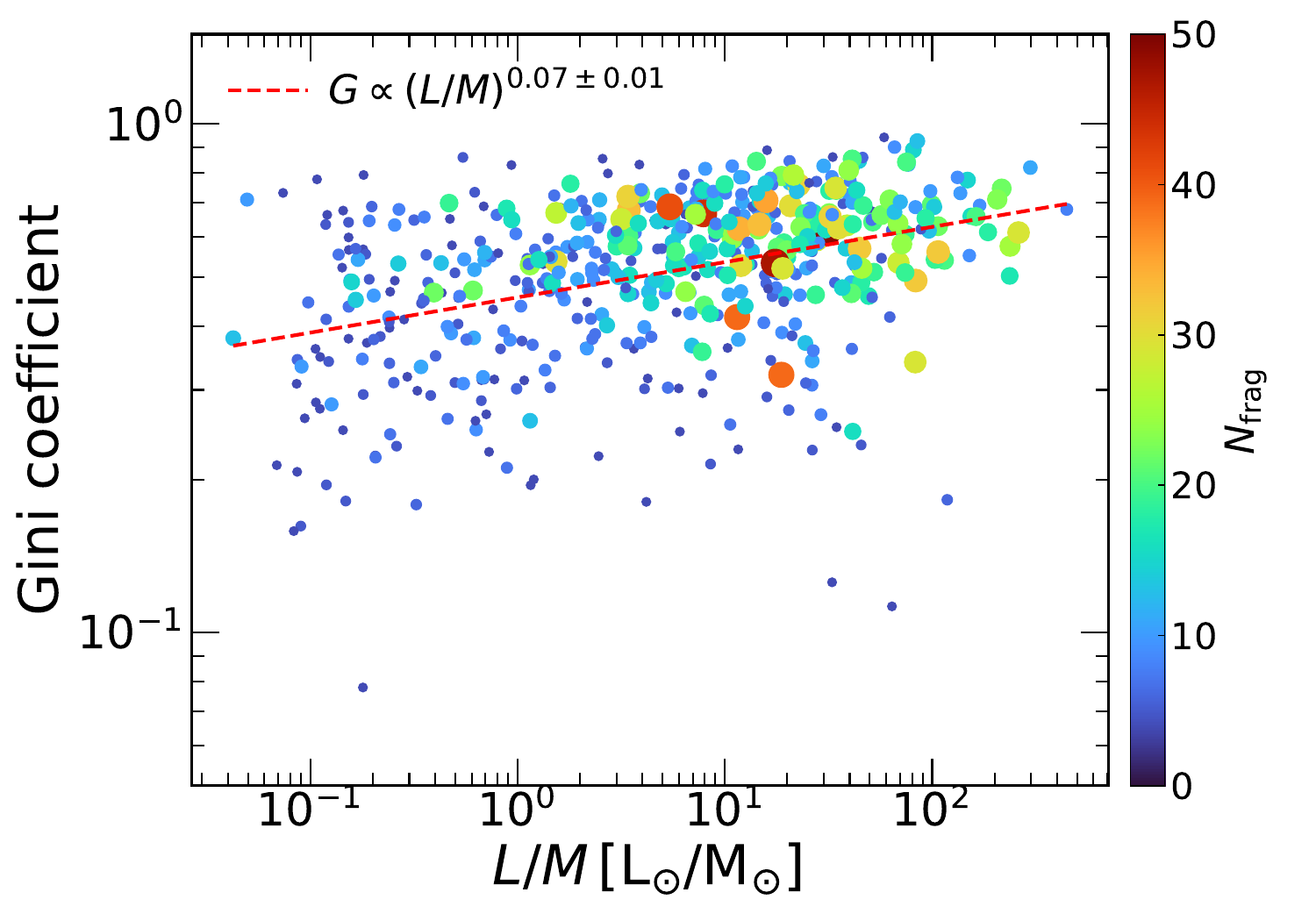}
\caption{Gini coefficient $G$ as a function of clump evolutionary stage $L/M$. The red dashed line shows the power-law fit $G \propto (L/M)^{0.07 \pm 0.01}$. Pearson's $r$ coefficient is $0.38$.}
\label{fig_gini_LM}
\end{figure}

\begin{figure}
\centering
\includegraphics[width=\linewidth]{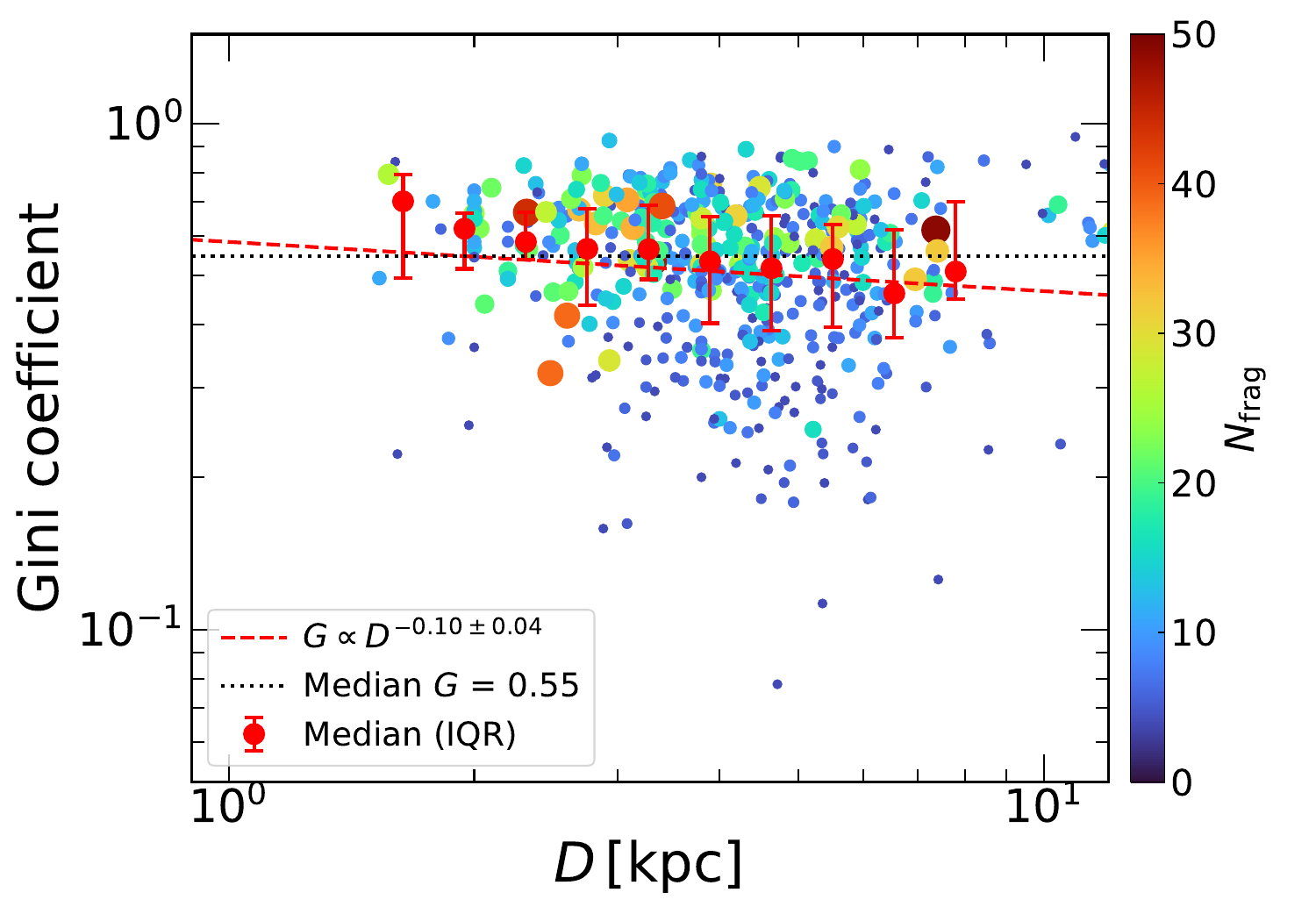}
\caption{Gini coefficient $G$ as a function of source distance $D$. A weak anticorrelation is found (Pearson $r = -0.100$, Spearman $\rho = -0.124$, $p = 0.005$), ruling out distance-dependent detection biases as the primary driver of the observed mass inequality trends. The red dashed line shows the power-law fit $G \propto D^{-0.10 \pm 0.04}$, and the black dotted line marks the median Gini value of $0.55$. Points are color-coded and sized by the number of cores per clump ($N_{\rm frag}$). Red circles with error bars represent the median and IQR (25th-75th percentiles) in bins of distance. The lack of correlation with distance confirms that the positive correlation between Gini and both $M_{\rm core,tot}$ (left panel of Figure~\ref{fig_gini_mass}) and $L/M$ (Figure~\ref{fig_gini_LM}) is not an artifact of distance-dependent sensitivity.}
\label{fig_gini_distance}
\end{figure}

\subsection{Core mass inequality: Gini coefficient analysis}
\label{sec_gini}

To quantify the inequality of core mass distributions within clumps, we compute the Gini coefficient for each clump with $N_{\rm frag} \ge 4$. The Gini coefficient ranges from 0 (perfect equality, all cores have identical masses) to 1 (perfect inequality, a single core dominates the total mass). The median Gini coefficient of our sample is $0.55 \pm 0.17$.

The left panel of Figure~\ref{fig_gini_mass} shows the adjusted Gini coefficient $G$ as a function of the total core mass $M_{\rm core,tot}$. A clear positive correlation is observed, with a power-law fit yielding (Pearson $r = 0.55$, $p \ll 10^{-5}$),
\begin{equation}
G = 3.61^{+0.10}_{-0.10} \times 10^{-1} \left( \frac{M_{\rm core,tot}}{1 \,\rm M_{\rm \odot}} \right)^{0.17 \pm 0.01} \, . 
\end{equation}
This indicates that clumps with larger total core masses exhibit more unequal core mass distributions.

To further investigate the origin of mass inequality, we examine the Gini coefficient as a function of the clump mass $M_{\rm clump}$. As shown in the right panel of Figure~\ref{fig_gini_mass}, no significant correlation is found (Pearson $r = -0.016$, Spearman $\rho \approx 0.0004$, $p = 0.99$), with a power-law fit $G \propto M_{\rm clump}^{-0.01 \pm 0.02}$ consistent with no dependence. This result, together with the strong positive correlation between Gini and $M_{\rm core,tot}$ (left panel of Figure~\ref{fig_gini_mass}), suggests that the mass inequality among cores is driven by the mass already assembled into cores, rather than the total clump mass reservoir.

Figure~\ref{fig_gini_LM} presents the adjusted Gini coefficient $G$ as a function of the clump evolutionary stage traced by $L/M$. A positive correlation is also found, with $G \propto (L/M)^{0.07 \pm 0.01}$ (Pearson $r = 0.38$, $p \ll 10^{-5}$). The median Gini increases from $0.49$ for clumps with $L/M \le 6.0$ to $0.62$ for $L/M > 6.0$ (Mann-Whitney $p = 4.23 \times 10^{-15}$). This demonstrates that the mass inequality among cores is not merely a mass-dependent effect but a genuine evolutionary trend: more evolved clumps have systematically more unequal core mass distributions. The results of the power-law fits are summarized in Table~\ref{tab_gini}. Additional analyses of the Gini coefficient as a function of fragmentation level ($N_{\rm frag}$) and clump surface density ($\Sigma_{\rm clump}$) are presented in Appendix~\ref{app_gini}.

Both the left panel of Figure~\ref{fig_gini_mass} and Figure~\ref{fig_gini_LM} demonstrate that mass inequality among cores increases with both total core mass and clump evolution. To rule out distance-dependent detection biases, we examine the Gini coefficient as a function of source distance. Figure~\ref{fig_gini_distance} shows a weak anticorrelation between Gini and distance (Pearson $r = -0.100$, Spearman $\rho = -0.124$, $p = 0.005$). The power-law fit yields $G \propto D^{-0.10 \pm 0.04}$, consistent with no dependence. This confirms that the observed increase of mass inequality with $M_{\rm core,tot}$ and $L/M$ is not an artifact of distance-dependent sensitivity, such as the inability to detect low-mass cores in distant clumps.

\begin{table}
\centering
\caption{Summary of Gini coefficient analysis.}
\begin{tabular}{lccc}
\hline
Relation & Exponent & Pearson $r$ & $p$ \\
\hline
Gini vs $M_{\rm core,tot}$ & $0.17 \pm 0.01$ & 0.55 & $<10^{-5}$ \\
Gini vs $L/M$ & $0.07 \pm 0.01$ & 0.38 & $<10^{-5}$ \\
\hline
\end{tabular}
\label{tab_gini}
\end{table}

\section{Discussion}
\label{discussion}

\subsection{The origin of the constant $f_{\rm core,max} \approx 0.38$}
\label{sec_discussion_f38}

The nearly constant mass fraction of the most massive core ($f_{\rm core,max} \approx 0.38$) across four orders of magnitude in total core mass is one of the most striking results of this study. This empirical relation implies a universal mass concentration efficiency in clump-fed core formation that is remarkably insensitive to the total mass budget.

A potential systematic uncertainty in our analysis is the possible contamination of free-free emission in the 1.38~mm continuum flux, which was not subtracted in the core mass derivation \citep{ALMAGAL+III+Coletta+2025A&A}. This contamination may lead to overestimates of core masses, particularly for the most massive cores in evolved clumps hosting HII regions. However, the contamination is expected to affect only a small fraction of the core population ($\sim 14\%$ of cores in the ALMAGAL catalog were found to be associated with radio continuum emission; see Paper III). Furthermore, our main findings on the constancy of $f_{\rm core,max}$ and the evolutionary growth of the Gini coefficient are based on the statistical trends across a large sample of 514 clumps, and are therefore unlikely to be significantly biased by this effect.

Our finding of a constant $f_{\rm core,max}$ is complementary to the analysis of core mass growth in \citet{ALMAGAL+III+Coletta+2025A&A}. They showed that the most massive cores grow in mass with evolution, while a population of lower-mass cores persists throughout the evolutionary sequence (see their Section~7). The constancy of $f_{\rm core,max}$ across four orders of magnitude in $M_{\rm core,tot}$ implies that this growth is self-similar: as the total mass in cores increases, the most massive core maintains a fixed share of the mass budget, suggesting that mass accretion is efficient and regulated. This is consistent with the competitive accretion scenario \citep{Bonnell+2001MNRAS}, where the most massive core deepens its gravitational potential and accretes preferentially, but our result suggests this process operates with a remarkably constant efficiency that is independent of the total mass available.

The CMF is observed to be significantly shallower than the IMF, suggesting that fragmentation alone does not directly produce the final IMF \citep{Louvet+ALMAIMF+2024A&A}. Our finding that $f_{\rm core,max}$ remains constant suggests that this shallowness may arise from a self-regulated process that limits the mass fraction of the most massive core.

Alternatively, a purely statistical interpretation is possible. If core masses follow a typical power-law distribution with slope $\alpha \sim -1.3$ to $-1.5$ (consistent with the CMF), the expected fraction of total mass in the maximum core for a sample of $N \sim 5-10$ cores is indeed in the range of $0.3-0.4$. The invariance of $f_{\rm core,max}$ may then reflect a self-similarity in the shape of the core mass distribution when normalized by total core mass. In the Orion Nebula Cluster region, \citet{Takemura+2021ApJ} found that the CMF peaks at $\sim 0.1 \, \rm M_\odot$ and that mass accretion from the surroundings is required to match the IMF, indicating that core growth continues after initial fragmentation.

In the competitive accretion scenario \citep{Bonnell+2001MNRAS}, the Bondi-Hoyle accretion rate $\dot{M} \propto M^2$ leads to preferential growth of already massive cores. This process may continue well beyond the initial fragmentation phase. \citet{Padoan+2025NatAs} show that Bondi-Hoyle accretion from the parental gas cloud supplies both mass and angular momentum to young stars on timescales of several million years after formation. However, this runaway growth must ultimately be limited by either gas exhaustion or feedback, and the observed $f_{\rm core,max} \approx 0.38$ may represent the asymptotic limit of this process.

The scatter in $f_{\rm core,max}$ spans approximately an order of magnitude (right panel of Figure~\ref{fig_max_core}). As shown in Figure~\ref{fig_max_core_lm_distance}, this scatter shows no significant correlation with either $L/M$ or source distance. The absence of correlation with $L/M$ is particularly noteworthy: if the scatter were driven by evolutionary differences, one would expect early-stage clumps (low $L/M$) to have lower $f_{\rm core,max}$ values (more equal mass distribution) and late-stage clumps (high $L/M$) to have higher $f_{\rm core,max}$ values (more dominated by a single core). The lack of such a trend suggests that the mass concentration efficiency is established early in the fragmentation process and remains largely unchanged thereafter. The absence of correlation with distance further confirms that the scatter is not a result of observational selection effects, such as the inability to detect low-mass cores in distant clumps. Instead, the scatter appears to be intrinsic to the fragmentation process, reflecting the stochastic nature of core formation and competitive accretion. Despite this intrinsic scatter, the median $f_{\rm core,max}$ remains remarkably constant at $\sim 0.37$, underscoring the universality of the mass concentration efficiency.

A definitive explanation for the constant $f_{\rm core,max} \approx 0.38$ remains elusive. Future work combining high-resolution numerical simulations of cluster formation with detailed radiative transfer will be required to determine whether this value emerges from first principles or reflects an as-yet-unidentified physical constraint on core growth. Having established the constancy of $f_{\rm core,max}$, we now turn to the broader evolution of mass inequality within core populations as quantified by the Gini coefficient.

\subsection{Implications for competitive accretion and evolutionary growth of inequality}

The Gini coefficient provides a holistic measure of mass inequality within core populations, complementing the $f_{\rm core,max}$ analysis that focuses solely on the most massive core. Our finding that Gini increases with both $M_{\rm core,tot}$ and $L/M$ offers two key insights.

First, the positive correlation with $M_{\rm core,tot}$ ($r = 0.55$) indicates that the ``mass budget'' of a clump directly influences the degree of inequality. The absence of a similar correlation with $M_{\rm clump}$ (Section~\ref{sec_gini}) further supports that it is the mass already assembled into cores, rather than the total clump mass reservoir, that drives the inequality. In clumps with larger total core masses, the distribution is more ``top-heavy'', meaning that a few massive cores claim a disproportionate share of the available mass. This aligns with the expectation of competitive accretion \citep{Bonnell+2001MNRAS}, where the most massive cores, once established, accrete more efficiently due to their deeper gravitational potential wells.

Second, the positive correlation with $L/M$ ($r = 0.38$) demonstrates that the increase in inequality is not a simple scaling with mass but reflects genuine evolutionary progression. Clumps at later evolutionary stages ($L/M > 6.0$) have significantly higher Gini coefficients (median $0.62$) than their less evolved counterparts ($L/M \le 6.0$, median $0.49$), with a Mann-Whitney $p = 4.23 \times 10^{-15}$. This suggests that inequality builds up over time, consistent with a scenario in which competitive accretion operates continuously throughout the clump's lifetime (see also \citet{Parker+2024ApJ} for a discussion of observational discriminants between competitive accretion and monolithic collapse).

Taken together, these results strengthen the case for a clump-fed, dynamically competitive mode of core assembly, where the final mass distribution of cores is not predetermined but emerges through time-dependent, scale-dependent accretion processes.

While our results strongly support the competitive accretion scenario, several open questions remain. Future work should investigate the connection between core mass inequality and the final stellar IMF, as well as the role of magnetic fields and turbulence in regulating the observed mass concentration efficiency. High-resolution simulations coupled with radiative transfer calculations will be essential to determine whether the max core mass fraction $f_{\rm core,max} \approx 0.38$ emerges from first principles or reflects an underlying physical constraint.

\section{Conclusions}
\label{summary}

We have analyzed the core mass distributions within ALMAGAL clumps using two complementary metrics: the mass fraction of the most massive core ($f_{\rm core,max}$) and the corrected Gini coefficient ($G$). Our analysis focuses on clumps with at least four cores ($N_{\rm frag} \ge 4$), comprising 514 clumps and 5,728 cores in total. Our main findings are:

(i) The mass fraction of the most massive core remains constant across a wide range of total core mass. The most massive core consistently accounts for $\sim 38\%$ of the total core mass across clumps spanning four orders of magnitude in $M_{\rm core,tot}$, with $f_{\rm core,max}$ showing no systematic trend with total core mass (Pearson $r = -0.02$). This result is consistent with the findings of \citet{ALMAGAL+III+Coletta+2025A&A} on the growth of massive cores, but extends their analysis by demonstrating that the mass concentration efficiency is universal and independent of the total mass budget. The distribution of $f_{\rm core,max}$ peaks around $0.3$--$0.4$ with a median of $0.37$ and a mean of $0.41$, demonstrating a universal mass concentration efficiency in clump-fed core accretion. The scatter in $f_{\rm core,max}$ is not correlated with evolutionary stage ($L/M$) or source distance, confirming that it is intrinsic to the fragmentation process rather than driven by evolutionary or observational biases.

(ii) Core mass inequality grows systematically with both total core mass and clump evolutionary stage. The Gini coefficient increases with $M_{\rm core,tot}$ as $G \propto M_{\rm core,tot}^{0.17 \pm 0.01}$ ($r = 0.55$, $p \ll 10^{-5}$) and with $L/M$ as $G \propto (L/M)^{0.07 \pm 0.01}$ ($r = 0.38$, $p \ll 10^{-5}$). Clumps at later evolutionary stages ($L/M > 6.0$) have significantly higher Gini values (median $0.62$) than their less evolved counterparts ($L/M \le 6.0$, median $0.49$; Mann-Whitney $p = 4.23 \times 10^{-15}$). Additional analyses show that the Gini coefficient also increases with fragmentation level ($G \propto N_{\rm frag}^{0.22 \pm 0.02}$, $r = 0.37$) and, more weakly, with clump surface density ($G \propto \Sigma_{\rm clump}^{0.18 \pm 0.02}$, $r = 0.32$).

These results provide strong support for the competitive accretion scenario. The observed growth of mass inequality with clump evolution demonstrates that core mass distributions are not static but evolve over time, consistent with a clump-fed, dynamically competitive mode of core assembly in which massive cores grow preferentially from a shared gas reservoir, while monolithic collapse models that assume predetermined core masses at formation cannot explain the observed evolutionary trend.


\section{Acknowledgements}

We thank the anonymous referee for their detailed and insightful comments that have improved the clarity of the paper. This research is supported by Guangxi Natural Science Foundation under Grant No.AD23026127, 2024GXNSFBA010436. ZJL acknowledges support by NSFC grant No.12563005, 12494571 and the National Key R\&D Program of China (Grant Nos. 2024YFA1611704, 2024YFA1611700). This research is also supported by Guangxi Qingmiao Talent Support Program, Guangxi Key Research and Development Program (Guike FN2504240040), Bagui Scholars Programme (W.X.-G., GXR-6BG2424001), Guangxi Talent Program (``Highland of Innovation Talents”). This paper makes use of the following ALMA data: ADS/JAO.ALMA\#2019.1.00195.L. ALMA is a partnership of ESO (representing its member states), NSF (USA) and NINS (Japan), together with NRC (Canada), NSTC and ASIAA (Taiwan), and KASI (Republic of Korea), in cooperation with the Republic of Chile. The Joint ALMA Observatory is operated by ESO, AUI/NRAO and NAOJ.


%




\appendix

\section{Distribution of the mass fraction of the most massive core}
\label{app_fmax}

\begin{figure}
\centering
\includegraphics[width=\linewidth]{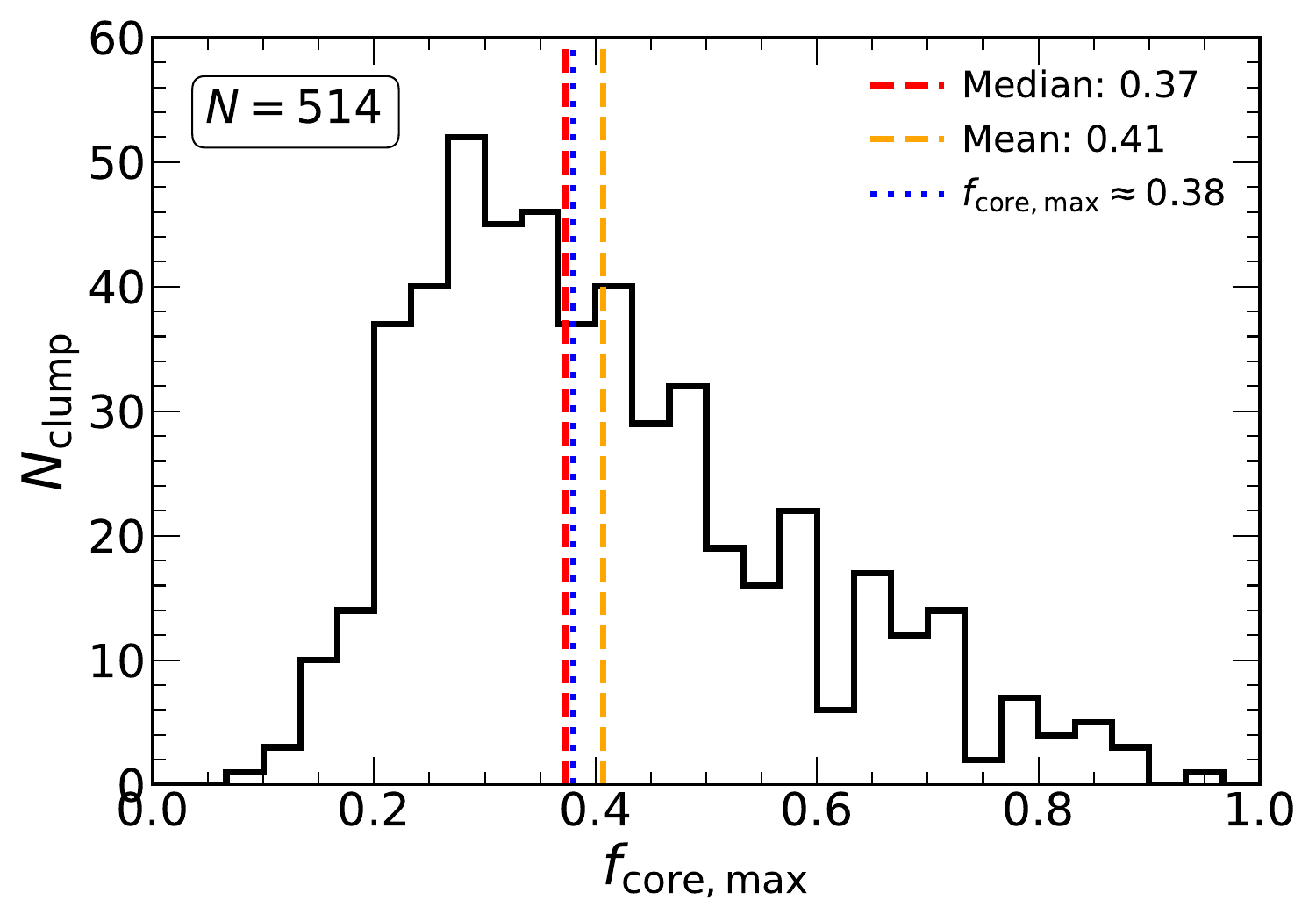}
\caption{Distribution of $f_{\rm core,max} = M_{\rm core,max} / M_{\rm core,tot}$ for clumps with $N_{\rm frag} \ge 4$. The red and orange dashed lines indicate the median (0.37) and mean (0.41), respectively. The blue dotted line marks $f_{\rm core,max} = 0.38$, the constant fraction from the power-law fit in the left panel of Figure~\ref{fig_max_core}. The distribution is asymmetric with a tail toward higher values.}
\label{fig_fmax_distribution}
\end{figure}

Figure~\ref{fig_fmax_distribution} shows the distribution of the mass fraction of the most massive core, $f_{\rm core,max}$, for the 514 clumps with $N_{\rm frag} \ge 4$. The distribution peaks around $f_{\rm core,max} \approx 0.3-0.4$, with a median of $0.37$ and a mean of $0.41$. The vertical dashed lines mark the median (red), mean (orange), and the constant fraction $f_{\rm core,max} \approx 0.38$ (blue dotted) found from the power-law fit in Figure~\ref{fig_max_core}. The distribution is asymmetric, with a tail extending to higher $f_{\rm core,max}$ values, indicating that while most clumps have $f_{\rm core,max} \sim 0.3-0.4$, a fraction of clumps exhibit more extreme mass concentration where a single core dominates the total core mass budget.

\section{Additional Gini coefficient analysis}
\label{app_gini}

In this appendix, we present two additional analyses of the Gini coefficient: as a function of fragmentation level ($N_{\rm frag}$) and as a function of clump surface density ($\Sigma_{\rm clump}$). These results complement the main analysis in Section~\ref{sec_gini}.

\subsection{Gini coefficient vs fragmentation level}

\begin{figure}
\centering
\includegraphics[width=\linewidth]{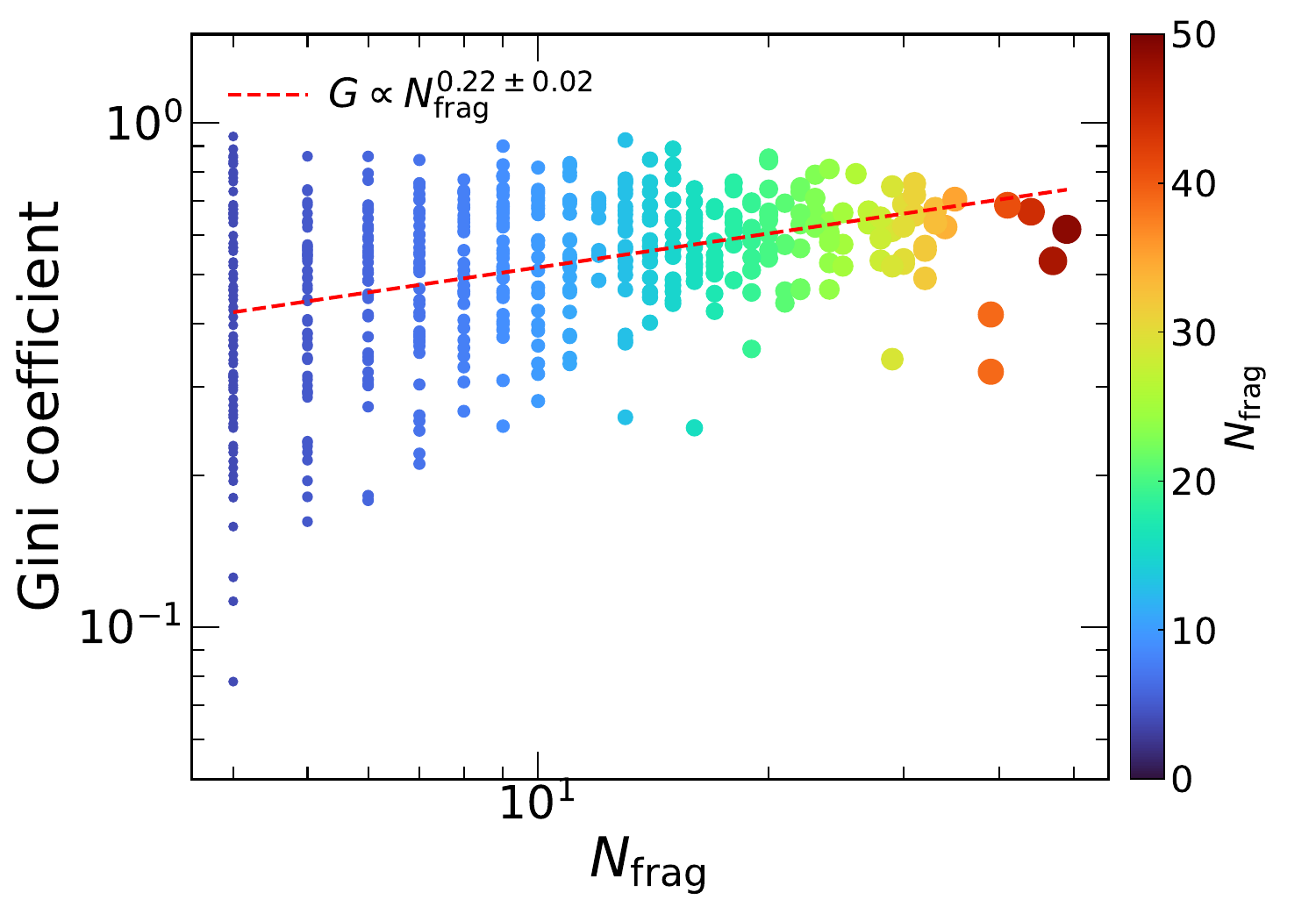}
\caption{Gini coefficient $G$ as a function of fragmentation level $N_{\rm frag}$. Points are color-coded and sized by $N_{\rm frag}$. The red dashed line shows the power-law fit $G \propto N_{\rm frag}^{0.22 \pm 0.02}$ (Pearson $r = 0.37$).}
\label{fig_gini_nfrag}
\end{figure}

Figure~\ref{fig_gini_nfrag} shows the adjusted Gini coefficient as a function of the number of cores per clump ($N_{\rm frag}$). A positive correlation is observed, with a power-law fit yielding
\begin{equation}
G = 0.309 \times N_{\rm frag}^{0.22 \pm 0.02} \, ,
\end{equation}
with Pearson correlation coefficient $r = 0.37$ ($p \ll 10^{-5}$). Clumps with $N_{\rm frag} > 9$ have higher Gini values (median $0.59$) than those with $N_{\rm frag} \le 9$ (median $0.49$; Mann-Whitney $p = 3.95 \times 10^{-11}$). This indicates that more fragmented clumps exhibit more unequal core mass distributions, consistent with the interpretation that fragmentation and mass inequality grow together during clump evolution.

\subsection{Gini coefficient vs clump surface density}

\begin{figure}
\centering
\includegraphics[width=\linewidth]{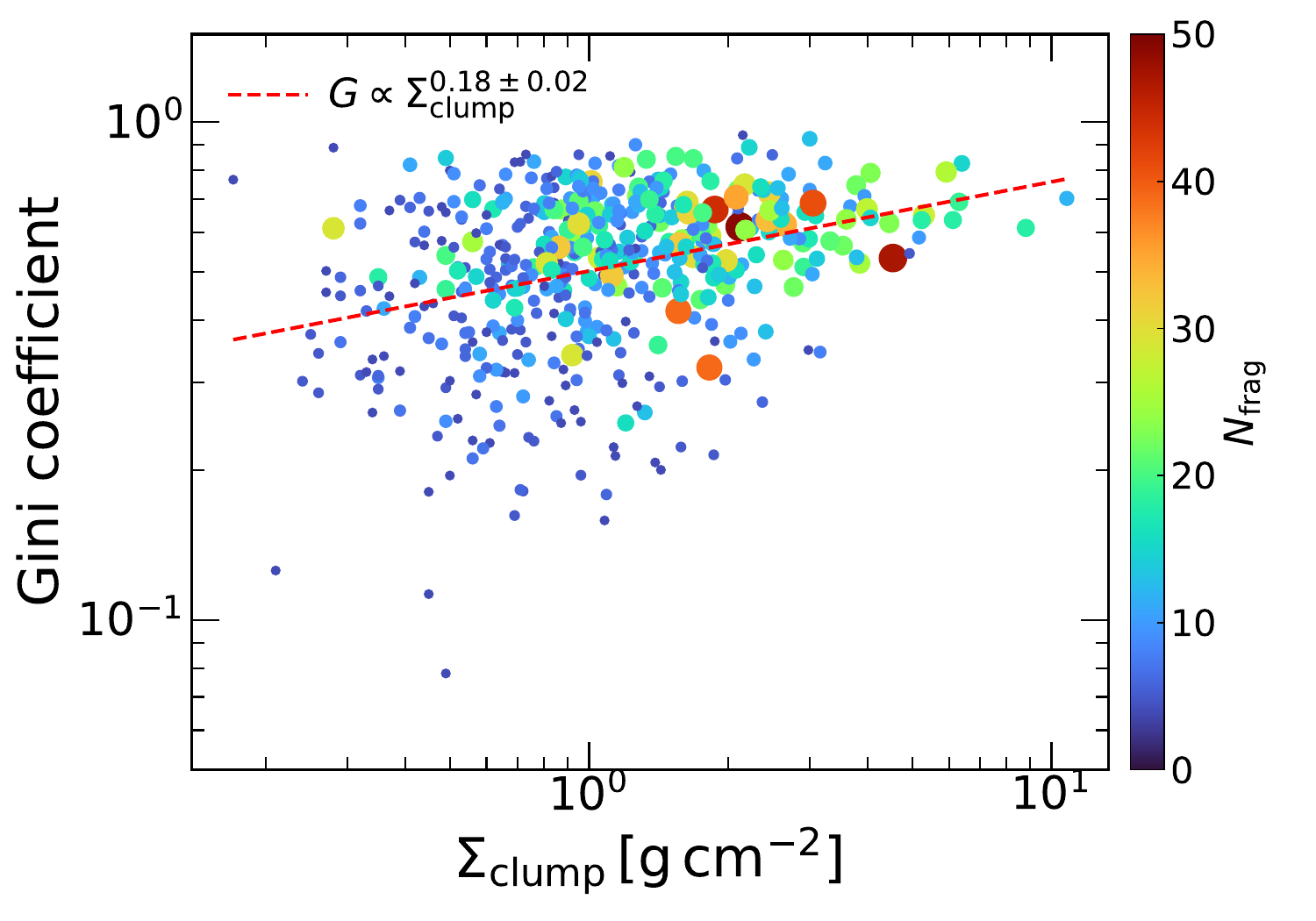}
\caption{Gini coefficient $G$ as a function of clump surface density $\Sigma_{\rm clump}$. Points are color-coded and sized by $N_{\rm frag}$. The red dashed line shows the power-law fit $G \propto \Sigma_{\rm clump}^{0.18 \pm 0.02}$ (Pearson $r = 0.32$).}
\label{fig_gini_sigma}
\end{figure}

Figure~\ref{fig_gini_sigma} presents the adjusted Gini coefficient as a function of clump surface density ($\Sigma_{\rm clump}$). A weak positive correlation is observed, with a power-law fit yielding
\begin{equation}
G = 0.502 \times \Sigma_{\rm clump}^{0.18 \pm 0.02} \, ,
\end{equation}
with Pearson correlation coefficient $r = 0.32$ ($p \ll 10^{-5}$). The correlation is weaker than those with $M_{\rm core,tot}$ ($r = 0.55$) and $L/M$ ($r = 0.38$), suggesting that surface density is not the primary driver of core mass inequality. This is consistent with the interpretation that the total mass available for core formation and the evolutionary stage of the clump are more fundamental factors.

\begin{table}
\centering
\caption{Summary of Gini coefficient correlations.}
\begin{tabular}{lccc}
\hline
Relation & Exponent & Pearson $r$ & $p$ \\
\hline
Gini vs $M_{\rm core,tot}$ & $0.17 \pm 0.01$ & 0.55 & $<10^{-5}$ \\
Gini vs $L/M$ & $0.07 \pm 0.01$ & 0.38 & $<10^{-5}$ \\
Gini vs $N_{\rm frag}$ & $0.22 \pm 0.02$ & 0.37 & $<10^{-5}$ \\
Gini vs $\Sigma_{\rm clump}$ & $0.18 \pm 0.02$ & 0.32 & $<10^{-5}$ \\
\hline
\end{tabular}
\label{tab_gini_appendix}
\end{table}

\subsection{Summary of Gini correlations}

Table~\ref{tab_gini_appendix} summarizes the correlation coefficients for all Gini analyses presented in this paper.


\bibliography{references}{}

\begin{thebibliography}{}
\expandafter\ifx\csname natexlab\endcsname\relax\def\natexlab#1{#1}\fi
\providecommand{\url}[1]{\href{#1}{#1}}
\providecommand{\dodoi}[1]{doi:~\href{http://doi.org/#1}{\nolinkurl{#1}}}
\providecommand{\doeprint}[1]{\href{http://ascl.net/#1}{\nolinkurl{http://ascl.net/#1}}}
\providecommand{\doarXiv}[1]{\href{https://arxiv.org/abs/#1}{\nolinkurl{https://arxiv.org/abs/#1}}}

\bibitem[{{Alves} {et~al.}(2007){Alves}, {Lombardi}, \& {Lada}}]{Alves+2007A&A}
{Alves}, J., {Lombardi}, M., \& {Lada}, C.~J. 2007, \aap, 462, L17, \dodoi{10.1051/0004-6361:20066389}

\bibitem[{{Beuther} {et~al.}(2018){Beuther}, {Mottram}, {Ahmadi}, {Bosco}, {Linz}, {Henning}, {Klaassen}, {Winters}, {Maud}, {Kuiper}, {Semenov}, {Gieser}, {Peters}, {Urquhart}, {Pudritz}, {Ragan}, {Feng}, {Keto}, {Leurini}, {Cesaroni}, {Beltran}, {Palau}, {S{\'a}nchez-Monge}, {Galvan-Madrid}, {Zhang}, {Schilke}, {Wyrowski}, {Johnston}, {Longmore}, {Lumsden}, {Hoare}, {Menten}, \& {Csengeri}}]{Beuther+2018A&A}
{Beuther}, H., {Mottram}, J.~C., {Ahmadi}, A., {et~al.} 2018, \aap, 617, A100, \dodoi{10.1051/0004-6361/201833021}

\bibitem[{{Bonnell} {et~al.}(2001){Bonnell}, {Bate}, {Clarke}, \& {Pringle}}]{Bonnell+2001MNRAS}
{Bonnell}, I.~A., {Bate}, M.~R., {Clarke}, C.~J., \& {Pringle}, J.~E. 2001, \mnras, 323, 785, \dodoi{10.1046/j.1365-8711.2001.04270.x}

\bibitem[{{Bonnell} {et~al.}(2004){Bonnell}, {Vine}, \& {Bate}}]{Bonnell+2004MNRAS}
{Bonnell}, I.~A., {Vine}, S.~G., \& {Bate}, M.~R. 2004, \mnras, 349, 735, \dodoi{10.1111/j.1365-2966.2004.07543.x}

\bibitem[{{Coletta} {et~al.}(2025){Coletta}, {Molinari}, {Schisano}, {Traficante}, {Elia}, {Benedettini}, {Mininni}, {Soler}, {S{\'a}nchez-Monge}, {Schilke}, {Battersby}, {Fuller}, {Beuther}, {Zhang}, {Beltr{\'a}n}, {Jones}, {Klessen}, {Walch}, {Fontani}, {Avison}, {Brogan}, {Clarke}, {Hatchfield}, {Hennebelle}, {Ho}, {Hunter}, {Johnston}, {Klaassen}, {Koch}, {Kuiper}, {Lis}, {Liu}, {Lumsden}, {Maruccia}, {M{\"o}ller}, {Moscadelli}, {Nucara}, {Rigby}, {Rygl}, {Sanhueza}, {van der Tak}, {Wells}, {Wyrowski}, {De Angelis}, {Liu}, {Ahmadi}, {Bronfman}, {Liu}, {Su}, {Tang}, {Testi}, \& {Zinnecker}}]{ALMAGAL+III+Coletta+2025A&A}
{Coletta}, A., {Molinari}, S., {Schisano}, E., {et~al.} 2025, \aap, 696, A151, \dodoi{10.1051/0004-6361/202452706}

\bibitem[{Deltas(2003)}]{Deltas2003}
Deltas, G. 2003, The Review of Economics and Statistics, 85, 226, \dodoi{10.1162/rest.2003.85.1.226}

\bibitem[{{Elia} {et~al.}(2017){Elia}, {Molinari}, {Schisano}, {Pestalozzi}, {Pezzuto}, {Merello}, {Noriega-Crespo}, {Moore}, {Russeil}, {Mottram}, {Paladini}, {Strafella}, {Benedettini}, {Bernard}, {Di Giorgio}, {Eden}, {Fukui}, {Plume}, {Bally}, {Martin}, {Ragan}, {Jaffa}, {Motte}, {Olmi}, {Schneider}, {Testi}, {Wyrowski}, {Zavagno}, {Calzoletti}, {Faustini}, {Natoli}, {Palmeirim}, {Piacentini}, {Piazzo}, {Pilbratt}, {Polychroni}, {Baldeschi}, {Beltr{\'a}n}, {Billot}, {Cambr{\'e}sy}, {Cesaroni}, {Garc{\'\i}a-Lario}, {Hoare}, {Huang}, {Joncas}, {Liu}, {Maiolo}, {Marsh}, {Maruccia}, {M{\`e}ge}, {Peretto}, {Rygl}, {Schilke}, {Thompson}, {Traficante}, {Umana}, {Veneziani}, {Ward-Thompson}, {Whitworth}, {Arab}, {Bandieramonte}, {Becciani}, {Brescia}, {Buemi}, {Bufano}, {Butora}, {Cavuoti}, {Costa}, {Fiorellino}, {Hajnal}, {Hayakawa}, {Kacsuk}, {Leto}, {Li Causi}, {Marchili}, {Martinavarro-Armengol}, {Mercurio}, {Molinaro}, {Riccio}, {Sano}, {Sciacca}, {Tachihara}, {Torii}, {Trigilio}, {Vitello}, \&
  {Yamamoto}}]{Elia+2017A&A}
{Elia}, D., {Molinari}, S., {Schisano}, E., {et~al.} 2017, \mnras, 471, 100, \dodoi{10.1093/mnras/stx1357}

\bibitem[{{Elia} {et~al.}(2021){Elia}, {Merello}, {Molinari}, {Schisano}, {Zavagno}, {Russeil}, {M{\`e}ge}, {Martin}, {Olmi}, {Pestalozzi}, {Plume}, {Ragan}, {Benedettini}, {Eden}, {Moore}, {Noriega-Crespo}, {Paladini}, {Palmeirim}, {Pezzuto}, {Pilbratt}, {Rygl}, {Schilke}, {Strafella}, {Tan}, {Traficante}, {Baldeschi}, {Bally}, {di Giorgio}, {Fiorellino}, {Liu}, {Piazzo}, \& {Polychroni}}]{Elia+2021A&A}
{Elia}, D., {Merello}, M., {Molinari}, S., {et~al.} 2021, \mnras, 504, 2742, \dodoi{10.1093/mnras/stab1038}

\bibitem[{{Elia} {et~al.}(2026){Elia}, {Coletta}, {Molinari}, {Schisano}, {Benedettini}, {S{\'a}nchez-Monge}, {Traficante}, {Mininni}, {Nucara}, {Pezzuto}, {Schilke}, {Soler}, {Avison}, {Beltr{\'a}n}, {Beuther}, {Clarke}, {Fuller}, {Klessen}, {Kuiper}, {Lebreuilly}, {Lis}, {M{\"o}ller}, {Moscadelli}, {Rigby}, {Sanhueza}, {van der Tak}, {Zhang}, {Rygl}, {Merello}, {Battersby}, {Ho}, {Klaassen}, {Koch}, {Allande}, {Bronfman}, {Fontani}, {Hennebelle}, {Jones}, {Liu}, {Stroud}, {Wells}, {Ahmadi}, {Brogan}, {De Angelis}, {Hunter}, {Johnston}, {Law}, {Liu}, {Liu}, {Maruccia}, {Pelkonen}, {Su}, {Tang}, {Testi}, {Walch}, {Zhang}, \& {Zinnecker}}]{ALMAGAL+V+Elia+2026A&A}
{Elia}, D., {Coletta}, A., {Molinari}, S., {et~al.} 2026, \aap, 705, A100, \dodoi{10.1051/0004-6361/202554764}

\bibitem[{{Gini}(1912)}]{Gini+1912amu.book}
{Gini}, C. 1912, {Variabilit{\`a} e mutabilit{\`a}}

\bibitem[{{Goyal} \& {Wang}(2022)}]{Goyal+2022ApJ}
{Goyal}, A.~V., \& {Wang}, S. 2022, \apj, 933, 162, \dodoi{10.3847/1538-4357/ac7562}

\bibitem[{{Kauffmann} \& {Pillai}(2010)}]{Kauffmann+2010ApJL}
{Kauffmann}, J., \& {Pillai}, T. 2010, \apjl, 723, L7, \dodoi{10.1088/2041-8205/723/1/L7}

\bibitem[{{Krumholz} {et~al.}(2019){Krumholz}, {McKee}, \& {Bland-Hawthorn}}]{Krumholz+2019ARA&A}
{Krumholz}, M.~R., {McKee}, C.~F., \& {Bland-Hawthorn}, J. 2019, \araa, 57, 227, \dodoi{10.1146/annurev-astro-091918-104430}

\bibitem[{{Louvet} {et~al.}(2024){Louvet}, {Sanhueza}, {Stutz}, {Men'shchikov}, {Motte}, {Galv{\'a}n-Madrid}, {Bontemps}, {Pouteau}, {Ginsburg}, {Csengeri}, {Di Francesco}, {Dell'Ova}, {Gonz{\'a}lez}, {Didelon}, {Braine}, {Cunningham}, {Thomasson}, {Lesaffre}, {Hennebelle}, {Bonfand}, {Gusdorf}, {{\'A}lvarez-Guti{\'e}rrez}, {Nony}, {Busquet}, {Olguin}, {Bronfman}, {Salinas}, {Fernandez-Lopez}, {Moraux}, {Liu}, {Lu}, {Huei-Ru}, {Towner}, {Valeille-Manet}, {Brouillet}, {Herpin}, {Lefloch}, {Baug}, {Maud}, {L{\'o}pez-Sepulcre}, \& {Svoboda}}]{Louvet+ALMAIMF+2024A&A}
{Louvet}, F., {Sanhueza}, P., {Stutz}, A., {et~al.} 2024, \aap, 690, A33, \dodoi{10.1051/0004-6361/202345986}

\bibitem[{{Lumsden} {et~al.}(2013){Lumsden}, {Hoare}, {Urquhart}, {Oudmaijer}, {Davies}, {Mottram}, {Cooper}, \& {Moore}}]{Lumsden+2013ApJS}
{Lumsden}, S.~L., {Hoare}, M.~G., {Urquhart}, J.~S., {et~al.} 2013, \apjs, 208, 11, \dodoi{10.1088/0067-0049/208/1/11}

\bibitem[{{McKee} \& {Tan}(2003)}]{McKee+2003ApJ}
{McKee}, C.~F., \& {Tan}, J.~C. 2003, \apj, 585, 850, \dodoi{10.1086/346149}

\bibitem[{{Molinari} {et~al.}(2016{\natexlab{a}}){Molinari}, {Merello}, {Elia}, {Cesaroni}, {Testi}, \& {Robitaille}}]{Molinari+2016aApJ}
{Molinari}, S., {Merello}, M., {Elia}, D., {et~al.} 2016{\natexlab{a}}, \apjl, 826, L8, \dodoi{10.3847/2041-8205/826/1/L8}

\bibitem[{{Molinari} {et~al.}(2010){Molinari}, {Swinyard}, {Bally}, {Barlow}, {Bernard}, {Martin}, {Moore}, {Noriega-Crespo}, {Plume}, {Testi}, {Zavagno}, {Abergel}, {Ali}, {Anderson}, {Andr{\'e}}, {Baluteau}, {Battersby}, {Beltr{\'a}n}, {Benedettini}, {Billot}, {Blommaert}, {Bontemps}, {Boulanger}, {Brand}, {Brunt}, {Burton}, {Calzoletti}, {Carey}, {Caselli}, {Cesaroni}, {Cernicharo}, {Chakrabarti}, {Chrysostomou}, {Cohen}, {Compiegne}, {de Bernardis}, {de Gasperis}, {di Giorgio}, {Elia}, {Faustini}, {Flagey}, {Fukui}, {Fuller}, {Ganga}, {Garcia-Lario}, {Glenn}, {Goldsmith}, {Griffin}, {Hoare}, {Huang}, {Ikhenaode}, {Joblin}, {Joncas}, {Juvela}, {Kirk}, {Lagache}, {Li}, {Lim}, {Lord}, {Marengo}, {Marshall}, {Masi}, {Massi}, {Matsuura}, {Minier}, {Miville-Desch{\^e}nes}, {Montier}, {Morgan}, {Motte}, {Mottram}, {M{\"u}ller}, {Natoli}, {Neves}, {Olmi}, {Paladini}, {Paradis}, {Parsons}, {Peretto}, {Pestalozzi}, {Pezzuto}, {Piacentini}, {Piazzo}, {Polychroni}, {Pomar{\`e}s}, {Popescu}, {Reach}, {Ristorcelli},
  {Robitaille}, {Robitaille}, {Rod{\'o}n}, {Roy}, {Royer}, {Russeil}, {Saraceno}, {Sauvage}, {Schilke}, {Schisano}, {Schneider}, {Schuller}, {Schulz}, {Sibthorpe}, {Smith}, {Smith}, {Spinoglio}, {Stamatellos}, {Strafella}, {Stringfellow}, {Sturm}, {Taylor}, {Thompson}, {Traficante}, {Tuffs}, {Umana}, {Valenziano}, {Vavrek}, {Veneziani}, {Viti}, {Waelkens}, {Ward-Thompson}, {White}, {Wilcock}, {Wyrowski}, {Yorke}, \& {Zhang}}]{Molinari+2010A&A}
{Molinari}, S., {Swinyard}, B., {Bally}, J., {et~al.} 2010, \aap, 518, L100, \dodoi{10.1051/0004-6361/201014659}

\bibitem[{{Molinari} {et~al.}(2016{\natexlab{b}}){Molinari}, {Schisano}, {Elia}, {Pestalozzi}, {Traficante}, {Pezzuto}, {Swinyard}, {Noriega-Crespo}, {Bally}, {Moore}, {Plume}, {Zavagno}, {di Giorgio}, {Liu}, {Pilbratt}, {Mottram}, {Russeil}, {Piazzo}, {Veneziani}, {Benedettini}, {Calzoletti}, {Faustini}, {Natoli}, {Piacentini}, {Merello}, {Palmese}, {Del Grande}, {Polychroni}, {Rygl}, {Polenta}, {Barlow}, {Bernard}, {Martin}, {Testi}, {Ali}, {Andr{\'e}}, {Beltr{\'a}n}, {Billot}, {Carey}, {Cesaroni}, {Compi{\`e}gne}, {Eden}, {Fukui}, {Garcia-Lario}, {Hoare}, {Huang}, {Joncas}, {Lim}, {Lord}, {Martinavarro-Armengol}, {Motte}, {Paladini}, {Paradis}, {Peretto}, {Robitaille}, {Schilke}, {Schneider}, {Schulz}, {Sibthorpe}, {Strafella}, {Thompson}, {Umana}, {Ward-Thompson}, \& {Wyrowski}}]{Molinari+2016bA&A}
{Molinari}, S., {Schisano}, E., {Elia}, D., {et~al.} 2016{\natexlab{b}}, \aap, 591, A149, \dodoi{10.1051/0004-6361/201526380}

\bibitem[{{Molinari} {et~al.}(2025){Molinari}, {Schilke}, {Battersby}, {Ho}, {S{\'a}nchez-Monge}, {Traficante}, {Jones}, {Beltr{\'a}n}, {Beuther}, {Fuller}, {Zhang}, {Klessen}, {Walch}, {Tang}, {Benedettini}, {Elia}, {Coletta}, {Mininni}, {Schisano}, {Avison}, {Law}, {Nucara}, {Soler}, {Stroud}, {Wallace}, {Wells}, {Ahmadi}, {Brogan}, {Hunter}, {Liu}, {Pezzuto}, {Su}, {Zimmermann}, {Zhang}, {Wyrowski}, {De Angelis}, {Liu}, {Clarke}, {Fontani}, {Klaassen}, {Koch}, {Johnston}, {Lebreuilly}, {Liu}, {Lumsden}, {Moeller}, {Moscadelli}, {Kuiper}, {Lis}, {Peretto}, {Pfalzner}, {Rigby}, {Sanhueza}, {Rygl}, {van der Tak}, {Zinnecker}, {Amaral}, {Bally}, {Bronfman}, {Cesaroni}, {Goh}, {Hoare}, {Hatchfield}, {Hennebelle}, {Henning}, {Kim}, {Kim}, {Maud}, {Merello}, {Nakamura}, {Plume}, {Qin}, {Svoboda}, {Testi}, {Veena}, \& {Walker}}]{ALMAGAL+I+Molinari+2025A&A}
{Molinari}, S., {Schilke}, P., {Battersby}, C., {et~al.} 2025, \aap, 696, A149, \dodoi{10.1051/0004-6361/202452702}

\bibitem[{{Motte} {et~al.}(1998){Motte}, {Andre}, \& {Neri}}]{Motte+1998A&A}
{Motte}, F., {Andre}, P., \& {Neri}, R. 1998, \aap, 336, 150

\bibitem[{{Padoan} {et~al.}(2025){Padoan}, {Pan}, {Pelkonen}, {Haugb{\o}lle}, \& {Nordlund}}]{Padoan+2025NatAs}
{Padoan}, P., {Pan}, L., {Pelkonen}, V.-M., {Haugb{\o}lle}, T., \& {Nordlund}, {\AA}. 2025, Nature Astronomy, 9, 862, \dodoi{10.1038/s41550-025-02529-3}

\bibitem[{{Parker} {et~al.}(2024){Parker}, {Pinson}, {Alcock}, \& {Dale}}]{Parker+2024ApJ}
{Parker}, R.~J., {Pinson}, E.~J., {Alcock}, H.~L., \& {Dale}, J.~E. 2024, \apj, 974, 8, \dodoi{10.3847/1538-4357/ad6c48}

\bibitem[{{Pelkonen} {et~al.}(2021){Pelkonen}, {Padoan}, {Haugb{\o}lle}, \& {Nordlund}}]{Pelkonen+2021MNRAS}
{Pelkonen}, V.-M., {Padoan}, P., {Haugb{\o}lle}, T., \& {Nordlund}, {\r{A}}. 2021, \mnras, 504, 1219, \dodoi{10.1093/mnras/stab844}

\bibitem[{{S{\'a}nchez-Monge} {et~al.}(2013){S{\'a}nchez-Monge}, {Cesaroni}, {Beltr{\'a}n}, {Kumar}, {Stanke}, {Zinnecker}, {Etoka}, {Galli}, {Hummel}, {Moscadelli}, {Preibisch}, {Ratzka}, {van der Tak}, {Vig}, {Walmsley}, \& {Wang}}]{SanchezMonge+2013A&A}
{S{\'a}nchez-Monge}, {\'A}., {Cesaroni}, R., {Beltr{\'a}n}, M.~T., {et~al.} 2013, \aap, 552, L10, \dodoi{10.1051/0004-6361/201321134}

\bibitem[{{Sanhueza} {et~al.}(2019){Sanhueza}, {Contreras}, {Wu}, {Jackson}, {Guzm{\'a}n}, {Zhang}, {Li}, {Lu}, {Silva}, {Izumi}, {Liu}, {Miura}, {Tatematsu}, {Sakai}, {Beuther}, {Garay}, {Ohashi}, {Saito}, {Nakamura}, {Saigo}, {Veena}, {Nguyen-Luong}, \& {Tafoya}}]{Sanhueza+2019ApJ}
{Sanhueza}, P., {Contreras}, Y., {Wu}, B., {et~al.} 2019, \apj, 886, 102, \dodoi{10.3847/1538-4357/ab45e9}

\bibitem[{{Schisano} {et~al.}(2026){Schisano}, {Molinari}, {Coletta}, {Elia}, {Schilke}, {Traficante}, {S{\'a}nchez-Monge}, {Beuther}, {Benedettini}, {Mininni}, {Klessen}, {Soler}, {Nucara}, {Pezzuto}, {van der Tak}, {Hennebelle}, {Beltr{\'a}n}, {Moscadelli}, {Rygl}, {Sanhueza}, {Koch}, {Lis}, {Kuiper}, {Fuller}, {Avison}, {Bronfman}, {Lebreuilly}, {M{\"o}ller}, {Liu}, {Pelkonen}, {Testi}, {Zhang}, {Zhang}, {Ahmadi}, {Allande}, {Battersby}, {Wallace}, {Brogan}, {Clarke}, {De Angelis}, {Fontani}, {Ho}, {Hunter}, {Jones}, {Johnston}, {Klaassen}, {Liu}, {Liu}, {Maruccia}, {Rigby}, {Su}, {Tang}, {Walch}, \& {Zinnecker}}]{ALMAGAL+VI+Schisano+2026A&A}
{Schisano}, E., {Molinari}, S., {Coletta}, A., {et~al.} 2026, \aap, 707, A221, \dodoi{10.1051/0004-6361/202555619}

\bibitem[{{Takemura} {et~al.}(2021){Takemura}, {Nakamura}, {Kong}, {Arce}, {Carpenter}, {Ossenkopf-Okada}, {Klessen}, {Sanhueza}, {Shimajiri}, {Tsukagoshi}, {Kawabe}, {Ishii}, {Dobashi}, {Shimoikura}, {Goldsmith}, {S{\'a}nchez-Monge}, {Kauffmann}, {Pillai}, {Padoan}, {Ginsberg}, {Smith}, {Bally}, {Mairs}, {Pineda}, {Lis}, {Burkhart}, {Schilke}, {Chen}, {Isella}, {Friesen}, {Goodman}, \& {Harper}}]{Takemura+2021ApJ}
{Takemura}, H., {Nakamura}, F., {Kong}, S., {et~al.} 2021, \apjl, 910, L6, \dodoi{10.3847/2041-8213/abe7dd}

\bibitem[{{Tan} {et~al.}(2014){Tan}, {Beltr{\'a}n}, {Caselli}, {Fontani}, {Fuente}, {Krumholz}, {McKee}, \& {Stolte}}]{Tan+2014PPV}
{Tan}, J.~C., {Beltr{\'a}n}, M.~T., {Caselli}, P., {et~al.} 2014, in Protostars and Planets VI, ed. H.~{Beuther}, R.~S. {Klessen}, C.~P. {Dullemond}, \& T.~{Henning}, 149--172, \dodoi{10.2458/azu_uapress_9780816531240-ch007}

\bibitem[{{Traficante} {et~al.}(2023){Traficante}, {Jones}, {Avison}, {Fuller}, {Benedettini}, {Elia}, {Molinari}, {Peretto}, {Pezzuto}, {Pillai}, {Rygl}, {Schisano}, \& {Smith}}]{Traficante+2023MNRAS}
{Traficante}, A., {Jones}, B.~M., {Avison}, A., {et~al.} 2023, \mnras, 520, 2306, \dodoi{10.1093/mnras/stad272}

\bibitem[{{Xu} {et~al.}(2024){Xu}, {Wang}, {Liu}, {Tang}, {Evans}, {Palau}, {Morii}, {He}, {Sanhueza}, {Liu}, {Stutz}, {Zhang}, {Chen}, {Li}, {G{\'o}mez}, {V{\'a}zquez-Semadeni}, {Li}, {Mai}, {Lu}, {Liu}, {Chen}, {Li}, {Shi}, {Ren}, {Li}, {Garay}, {Bronfman}, {Dewangan}, {Juvela}, {Lee}, {Zhang}, {Yue}, {Wang}, {Ge}, {Jiao}, {Luo}, {Zhou}, {Tatematsu}, {Chibueze}, {Su}, {Sun}, {Ristorcelli}, \& {Toth}}]{Xu+2024ApJS}
{Xu}, F., {Wang}, K., {Liu}, T., {et~al.} 2024, \apjs, 270, 9, \dodoi{10.3847/1538-4365/acfee5}

\end{thebibliography}
\bibliographystyle{aasjournal}



\end{document}